\documentclass[english,showpacs,floatfix]{revtex4}
\usepackage[dvips]{graphicx}
\usepackage{longtable}
\usepackage{amsmath,amssymb}
\usepackage{dcolumn}

\usepackage{babel}
\usepackage[latin1]{inputenc}

\begin{document}

\title{The signals from the brane-world black Hole}

\author{Jianyong Shen}
\author{Bin Wang}
\email{wangb@fudan.edu.cn}
\affiliation{Department of Physics, Fudan University, Shanghai
200433, People's Republic of China }

\author{Ru-Keng Su}
\email{rksu@fudan.ac.cn}
\affiliation{China Center of Advanced Science and Technology (World
Laboratory) P.O. Box 8730, Beijing 100080, People's Republic of
China}\affiliation{Department of Physics, Fudan University, Shanghai
200433, People's Republic of China }


\begin{abstract}
We have studied the wave dynamics and the Hawking radiation for the scalar field as well as the brane-localized gravitational field
in the background of the braneworld black hole with tidal charge containing information of the extra dimension.
Comparing with the four-dimensional  black holes, we have observed the
signature of the tidal charge which presents the signals of the extra dimension both in the wave dynamics and the Hawking radiation.
\end{abstract}

\maketitle

\section{Introduction}
In the past years there has been growing interest on studying models
with extra dimensions in which the standard model fields are
confined on a 3-brane playing the role of our 4-dimensional world,
while gravity can propagate both on the brane and in the bulk
\cite{s1}\cite{randall}\cite{s3}. The extra dimensions need not be
compact and in particular it was shown that it is possible to
localize gravity on a 3-brane when there is one infinite extra
dimension \cite{randall}\cite{s3}. One of the striking consequences
of the theories with large extra dimensions is that the lowering of
the fundamental gravity scale allows the production of mini black
holes in the universe. Such mini black holes are centered on the
brane and may have been created in the early universe due to density
perturbations and phase transitions. Recently it was proposed that
such mini black holes may also be produced in particle collision
with the mass energy of TeV order or in the earth atmosphere due to
the high energy cosmic ray showers \cite{s4}\cite{s5}\cite{s6}. Once
produced, these black holes will go through a number of stages in
their lives \cite{s5}\cite{s6}\cite{s7}, namely: i) the balding
phase, where the black hole sheds the hair inherited from original
ordinary object; ii) the spin-down phase, where the black hole loses
its angular momentum; iii) the Schwarzschild quantum phase, where
the black hole's mass will be decreased due to quantum process and
finally iv) the Planck Phase. In the first phase, the black hole
emits mainly gravitational radiation, while in the second and the
third phases, the black hole will lose its energy mainly through the
emission of the Hawking radiation.

It was argued that during a certain time interval the evolution of
the perturbation around black hole is dominated by damped
single-frequency oscillation, called quasinormal modes (QNM), which
carries unique fingerprint of the black hole and is expected to be
detected through gravitational wave observations in the near future
(see reviews on this topic and references therein \cite{s8}).
Recently the QNM of a brane-world black hole has been studied in
[13]. The gravitational radiation of the mini black hole would
be a characteristic sound and can tell us the existence of such
black hole. Another possibility of observing signatures of this kind
of mini black hole exists in particle accelerator experiments, where
the spectrum of Hawking radiation emitted by these small black holes
can be detected \cite{s9}\cite{s10}\cite{s11}\cite{s12}\cite{park}.
Since these small black holes carry information of extra dimensions
and have different properties compared to ordinary 4-dimensional
black holes, these two tools of detecting mini black holes can help
to read the extra dimensions.

The motivation of the present paper is to study signals of tiny
black holes in the modern brane world scenarios[12]. Under the
assumptions of the theory, most standard matter fields are
brane-localized, therefore from the observational point of view it
is much more interesting to study the brane-localized modes in the
QNM and Hawking radiation. In the Hawking radiation study, it was
found that the emission on the brane is dominated compared to that
off the brane \cite{s10}. We will investigate brane world black
holes in the 4-dimensional background and study the QNM and the
emission of brane-localized Hawking radiation. By comparing the
properties of QNM and Hawking radiation to those in ordinary
4-dimensional black holes, we will argue that the dependence of
these properties on the extra dimension of spacetime could be read
if the spectrum of QNM and Hawking radiation are detected. In our
following study, we will employ the exact black hole solution to the
effective field equations on the brane, which is of the
Reissner-Nordstrom type given as \cite{Tdl}
\begin{equation} \label{e0}
ds_4^2  =  - f dt^2+f^{-1}dr^2+r^2 d\Omega ^2
\end{equation}
where $f=1-\frac{2M}{M_{4}^{2}r}+\frac{q}{\tilde{M}_P^2 r^2}$, and
$q$ is not the electric charge of the conventional
Reissner-Nordstrom (RN) metric, but the 'tidal charge' arising from
the projection onto the brane of the gravitational field in the
bulk. Thus $q$ contains the information of the extra dimension. For
$q>0$, this metric is a direct analogy to the Reissner-Nordstrom
solution with two horizons and both horizons lie inside the
Schwarzschild horizon $2M/M_4^2$. For $q<0$, the metric has only one
horizon given by $ r_ +   = \frac{M}{{M_4^2 }}(1 + \sqrt {1 -
\frac{{qM_4^4 }}{{M^2 \tilde M^4_p }}} )$, which is larger than the
Schwarzschild horizon. We will concentrate our attention on $q<0$,
since it was argued that this negative tidal charge is the
physically more natural case \cite{Tdl}.

Astrophysics limits the appearance of the Reissner-Nordstrom black
hole with macroscopic electric charge, while there is no exact
constraint on the emergence of the effective Reissner-Nordstrom
black hole with tidal charge, especially negative tidal charge.
However the tidal charge affects the geodesics and the gravitational
potential, so that its value should receive at least indirect limit
from observations, which needs careful study. On the other hand,
tidal charge contains the information of extra dimensions. It is of
great interest to investigate its influence on the gravitational
wave and Hawking radiation observations and whether it can leave us
the signature of the extra dimension.

This paper is organized as following. In Sec.II, we will investigate
the QNM of the braneworld black hole with tidal charge. We will
study the scalar as well as the gravitational perturbations in the
background of this black hole. In Sec.III, we will examine the
absorption and emission problems for the scalar and the
brane-localized graviton in the background of the braneworld black
hole. Our conclusions are summarized in Sec.IV. For simplicity we
will adopt the nature units in the following discussion.

\section{Quasi-Normal Modes}
The effective 4D field equation on the brane, which is induced by
the 5D Einstein equation in the bulk, is written as \cite{Z2}
\begin{equation} \label{e2}
R_{\mu \nu }  =  - E_{\mu \nu } ,\quad R_{\mu} ^{\mu}  = 0 = E_{\mu}
^{\mu},
\end{equation}
where $E_{\mu \nu}$ is the symmetric and tracefree limit on the
brane of projected bulk Weyl tensor $E_{[\mu, \nu]}= 0
=E^{\mu}_{\mu} $, if we only take the vacuum solution into account
and take a vanishing cosmological constant. The Bianchi Identity on
the brane requires the constraint
\begin{equation} \label{e3}
\nabla ^\mu  E_{\mu \nu }  = 0.
\end{equation}
As mentioned in \cite{maartens}, Eq.(\ref{e2}) and Eq.(\ref{e3})
form a closed system of equations on the brane for static solutions.
Dadhich et.al. \cite{Tdl} investigated one of the vacuum static
solutions on the brane and pointed out that Einstein-Maxwell
solutions in general relativity provides a vacuum static brane-world
solution, because both $E_{\mu \nu}$ from the influence of bulk and
the energy momentum tensor of electromagnetic field satisfy
Eq.(\ref{e2}) and Eq.(\ref{e3}) in spite of their different origins.
Hence the 4D metric on the brane is given by
\[
ds_4^2  =  - fdt^2  + f^{ - 1} dr^2  + r^2 d\Omega ^2,
\]
\begin{equation} \label{e4}
f = 1 - \frac{{2M}}{{M_p^2 }}\frac{1}{r} + \frac{q}{{\tilde M_p^2
}}\frac{1}{{r^2 }},
\end{equation}
where $q$ is a dimensionless tidal charge parameter. When $q > 0$,
this solution has two horizons and has similar properties to that of
the Reissner-Nordstrom solution. The new situation occurs when $q <
0$ and the black hole has only one horizon which is bigger than the
Schwarzschild horizon. So the negative tidal charge increases the
entropy and decreases the temperature of the black hole. With the
negative tidal charge the bulk influences tend to strengthen the
gravitational field. This opinion was supported by the perturbative
\cite{sasaki} and nonperturbative analysis \cite{maartens}. We will
concentrate our discussion on the brane world black hole with the
negative $q$. It is worthy to emphasis that although the 4D metric
on the brane is known as Eq.(\ref{e4}), the exact solution in the 5D
bulk has not been given yet and is still a challenging task to find.
Choosing negative $q$ (in the following we express the negative
tidal charge to be $Q$), for simplicity we rewrite the metric on the
4D brane
\begin{equation} \label{e5}
f = 1 - \frac{{2M}}{r} - \frac{Q}{{r^2 }} \equiv \frac{1}{{r^2
}}(r - r_ +  )(r - r_ -  ),
\end{equation}
where $r_+$ and $r_-$ are two roots of $f=0$ and
\begin{equation}\label{e6}
\begin{array}{l}
 r_ +   = M(1 + \sqrt {1 + \frac{Q}{{M^2 }}} ) \\
 r_ -   = M(1 - \sqrt {1 + \frac{Q}{{M^2 }}} ). \\
 \end{array}
\end{equation}
$r_+$ is the black hole horizon, while $r_{-}$ is negative and
without physical meaning. Now we start to consider the perturbations
around this black hole. For massless scalar perturbations, the wave
equation is given by the Klein-Gordon equation
\begin{equation} \label{e7}
\nabla ^\mu  \nabla _\mu  \phi  = 0,
\end{equation}
which can be reduced to
\begin{eqnarray} \label{e8}
 \frac{{\partial ^2 \Psi {}_l}}{{\partial r_* ^2 }} -
 \frac{{\partial ^2 \Psi _l }}{{\partial t^2 }} = V_s (r)\Psi _l,
\end{eqnarray}
where:
\begin{eqnarray} \label{addition}
 V_s (r) = f\left[ {\frac{{l(l + 1)}}{{r^2 }} + \frac{{f'}}{r}}
 \right] \nonumber
\end{eqnarray}
is the effective potential. In deriving the above equation, we have
used the variables separation $\phi = \sum\limits_l {\frac{{\Psi _l
(r,t)}}{r}Y_{lm} (\theta ,\varphi )}$ and introduced the tortoise
coordinate, given by
\begin{equation} \label{e9}
dr_*=dr/f \quad  \Rightarrow \quad  r_*  = r + \frac{{r_ +  ^2
}}{{r_ + - r_ - }}\ln (r - r_ + ) + \frac{{r_ -  ^2 }}{{r_ -   - r_
+ }}\ln (r - r_ -  ).
\end{equation}
\begin{figure}[ht]
\vspace*{0cm}
\begin{minipage}{0.3\textwidth}
\resizebox{1.1\linewidth}{!}{\includegraphics*{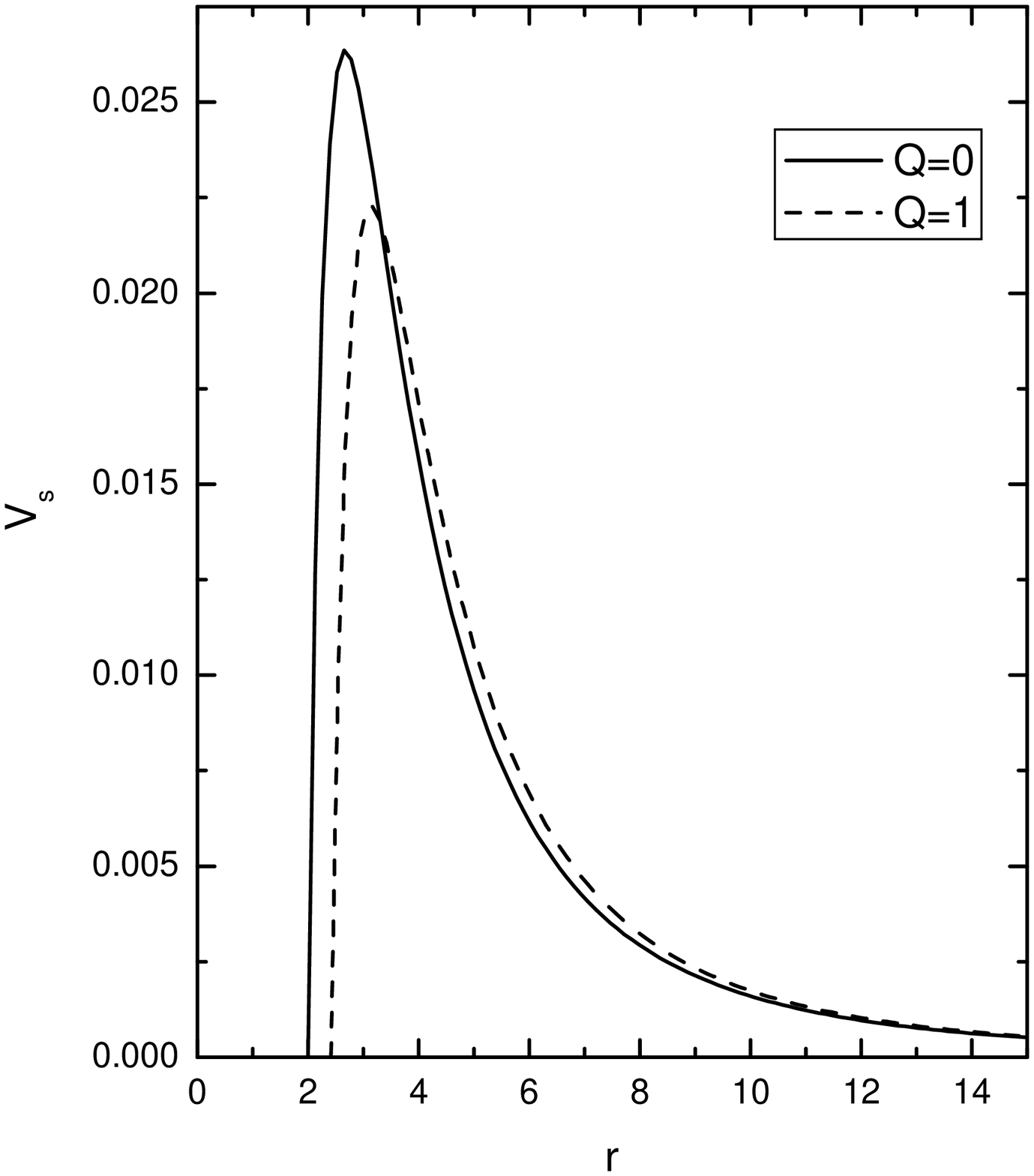}}
\nonumber
\end{minipage}\nonumber
\begin{minipage}{0.3\textwidth}
\vspace*{0.0cm}
\resizebox{1.1\linewidth}{!}{\includegraphics*{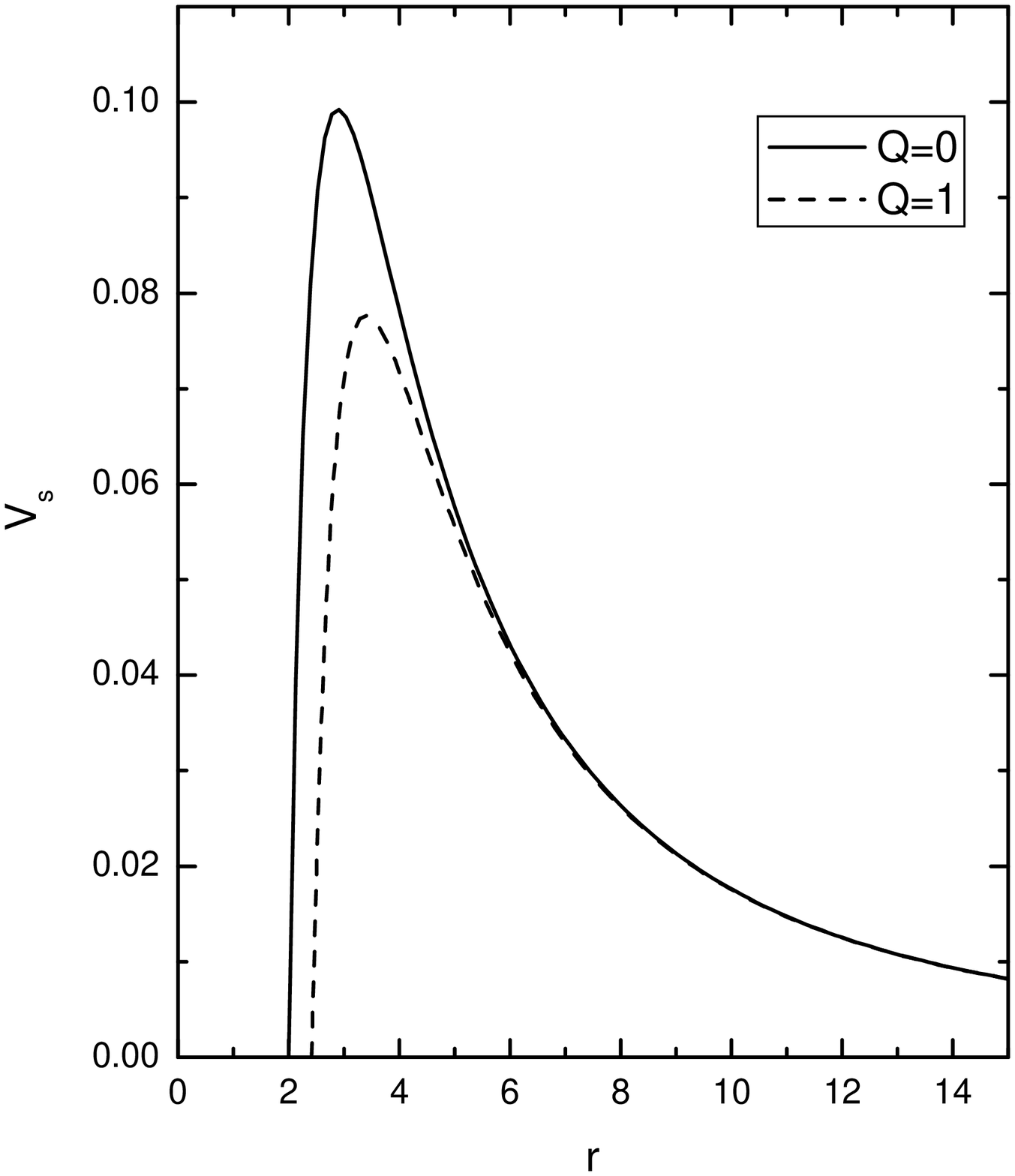}}
\nonumber
\end{minipage} \nonumber
\begin{minipage}{0.3\textwidth}
\vspace*{0.0cm}
\resizebox{1.1\linewidth}{!}{\includegraphics*{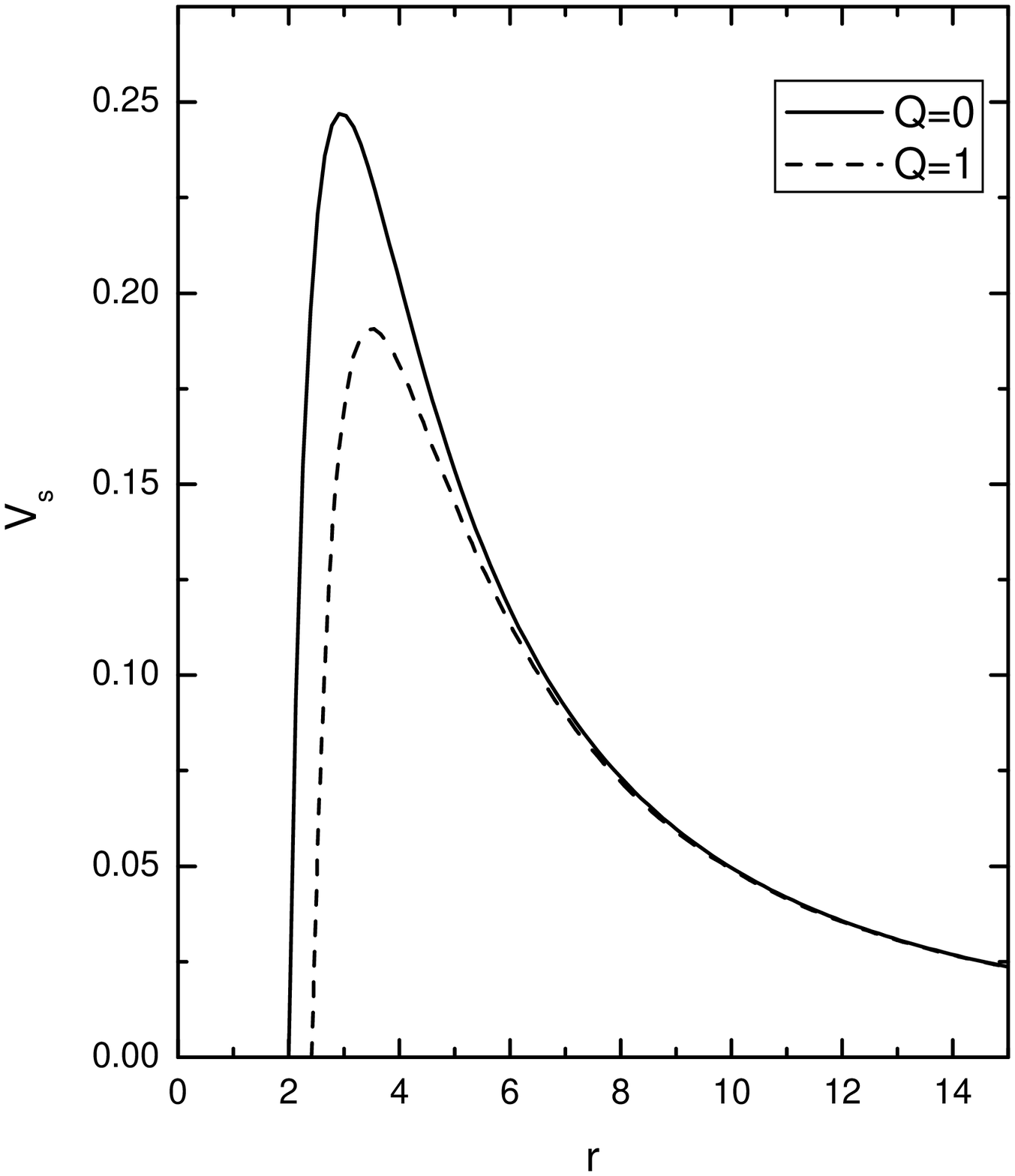}}
\nonumber
\end{minipage} \nonumber
\caption{{The behavior of the effective potential of the massless
scalar wave equation. We have taken $M=1$, $l=0,1,2$ respectively
from the left to the right.}}\label{f1}
\end{figure}
The behaviors of the effective potential $V_s$ are shown in
Fig.\ref{f1}. We see that just outside the black hole, there is a
potential barrier. This barrier increases with the increase of the
angular index $l$. While for the same $l$, the appearance of the
negative tidal charge as a sequence of strengthening the
gravitational field by the bulk effects suppressed the potential
barrier.

\begin{figure}[ht]
\vspace*{0cm}
\begin{minipage}{0.4\textwidth}
\resizebox{0.9\linewidth}{!}{\includegraphics*{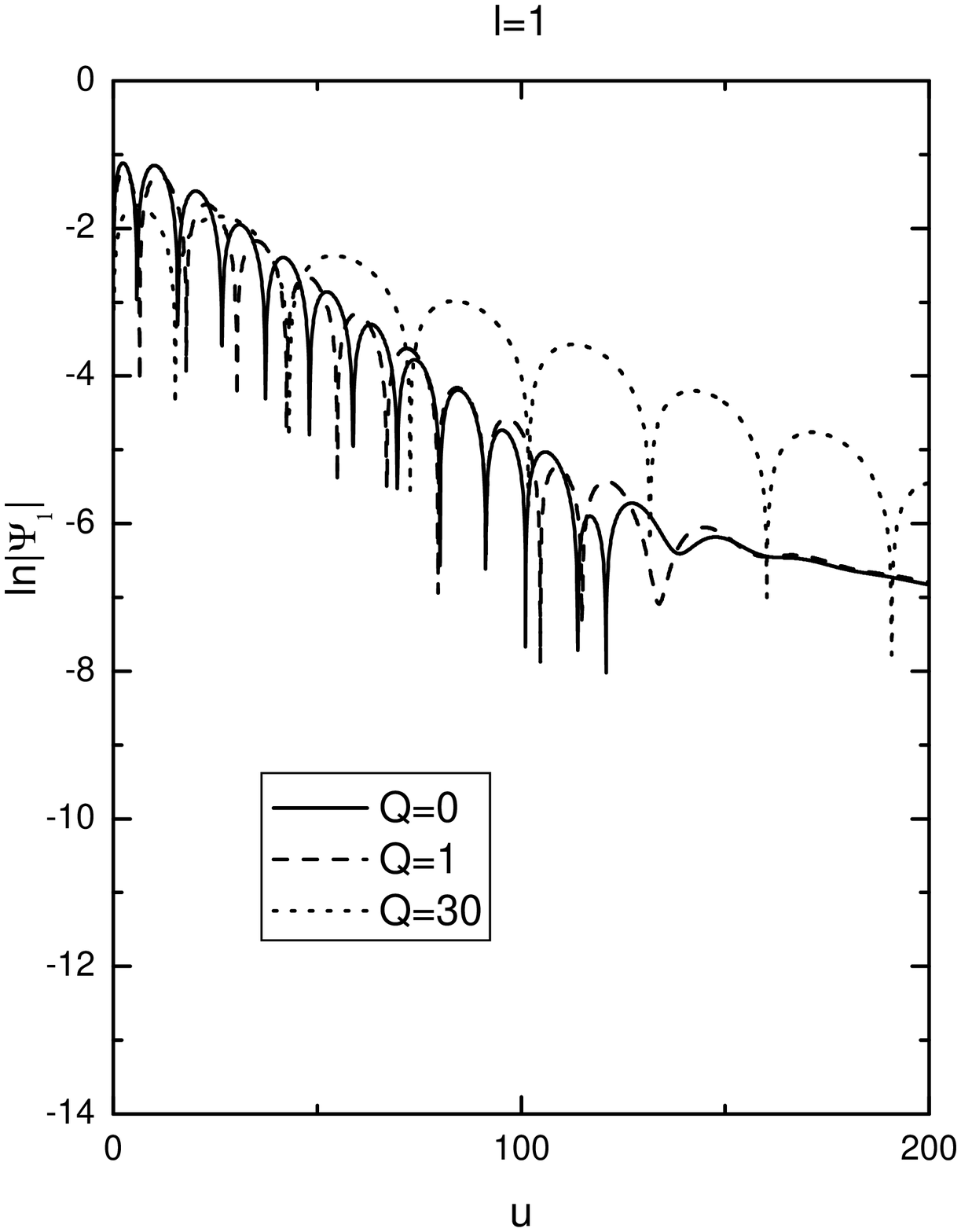}}
\nonumber
\end{minipage}\nonumber
\begin{minipage}{0.4\textwidth}
\vspace*{0.0cm}
\resizebox{0.9\linewidth}{!}{\includegraphics*{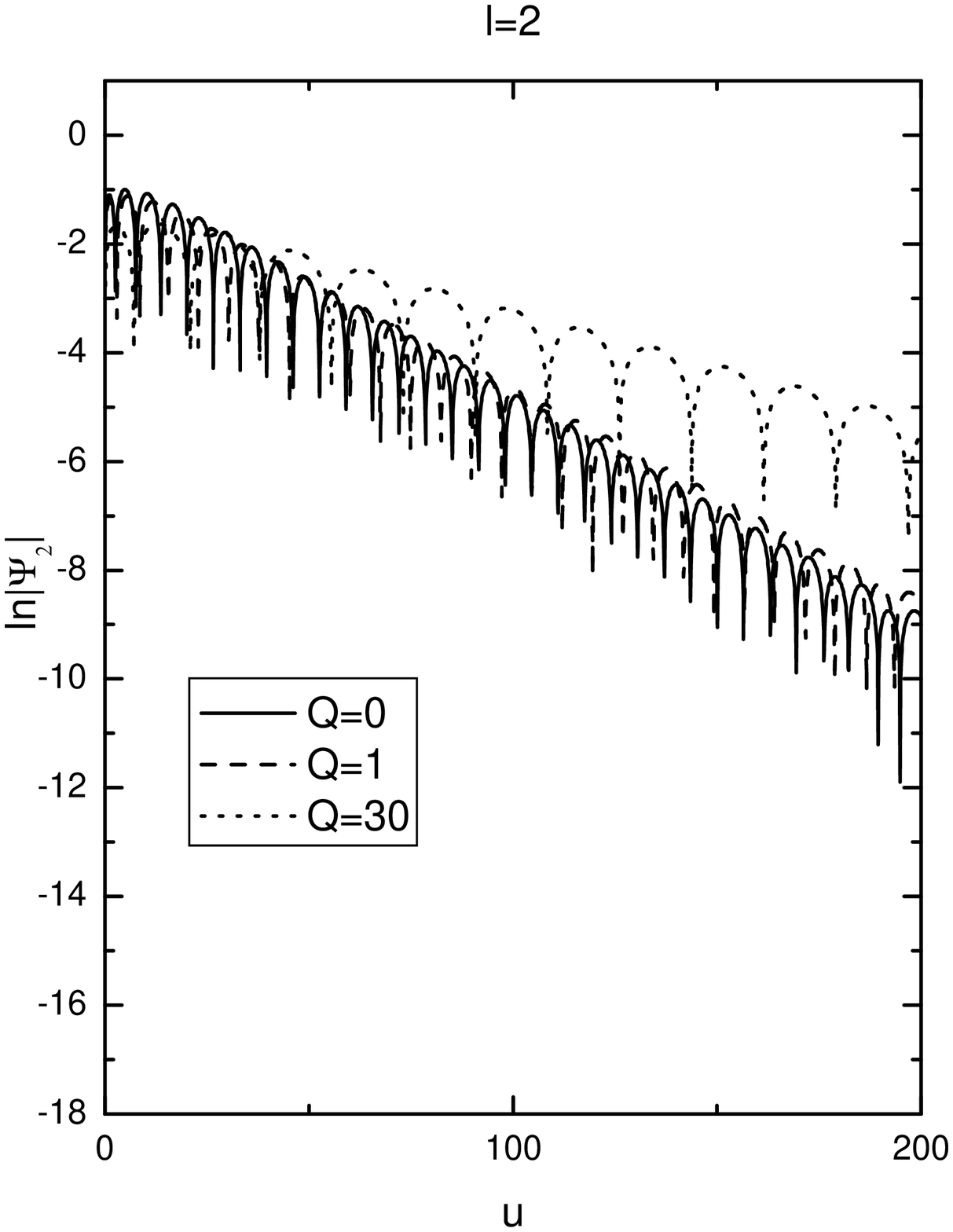}}
\nonumber
\end{minipage} \nonumber
\caption{{The QNMs of the Schwarzschild black hole and the
brane-world black hole. We have taken the mass $M=1$.}}\label{f8}
\end{figure}

Due to the potential barrier out of the black hole, the massless
scalar perturbation will experience the quasinormal ringing. Using
the null coordinates $u=t-r_*$ and $v=t+r_*$, we can rewrite Eq(8)
as
\begin{equation} \label{e10}
-4\frac{\partial^2 \Psi_l}{\partial u\partial v}=V(r)\Psi_l,
\end{equation}
where $r$ should be obtained by the inverse relation $r_*(r)=(v-u)/2$. We can numerically solve this equation
by the finite difference method
\begin{equation} \label{e11}
\Psi_N=\Psi_E+\Psi_W-\Psi_S-\delta u\delta v V_s (\frac{v_N+v_W-u_N-u_E}{4})\frac{\Psi_E+\Psi_W}{8},
\end{equation}
where points $N,S,E,W$ form a null rectangle with the relative position as $N:(u+\delta u,v+\delta v),
W:(u+\delta u, v), E:(u, v+\delta v)$ and $S:(u,v)$.
Employing this numerical method proposed in \cite{price}, we can
observe the object picture of the quasinormal ringing and obtain the
frequencies of the fundamental mode. Fig.\ref{f8} shows the massless
scalar wave outside the negative tidal charge black hole and Table.I
presents us the frequencies. For the negative tidal charge, we
observed that when the $Q$ increases, the massless scalar
perturbation will have less oscillations and decay more slowly. This
behavior of the wave evolution is monotone with the increase of $Q$
if the tidal charge $q=-Q$ is negative. For comparison, we have also
calculated the QNM for positive charged black hole, the result is
similar to that of the RN black hole disclosed in
\cite{star}\cite{wangb1}\cite{wangb2}. Unlike the negative charged
case, for the positive charge, the quasinormal ringing decays faster
than the case when $Q=0$. The QNM of the black hole for the $Q=0$
case, which is the 4D Schwarzschild black hole, serves as a border
separating the quasinormal ringing behaviors of the positive and
negative charged black hole.

\vspace{0.5cm}
\begin{center}

$\omega = \omega_R+ i \omega_I$

\begin{tabular}{l|l|l|l|r} \hline
 & $l=1$ & $l=2$ \\ \hline \hline
RN. $Q=-0.5$&   $0.32 - i 0.087$  &   $0.53 - i 0.086$  \\
Sch. $Q=0$  &   $0.29 - i 0.085$   &  $0.48 - i 0.084$       \\
Tdl. $Q=1$  &   $0.26 - i 0.079$   &  $0.42 - i 0.079$   \\
Tdl. $Q=30$  &   $0.053 - i 0.010$   &  $0.088 - i 0.010$   \\
\hline
\end{tabular}

\vspace{0.1cm} \large{Table I}

\vspace{0.1cm} {\it Quasinormal frequencies of the massless scalar
field in the Schwarzchild and tidal charge black hole
backgrounds. }

\end{center}
\vspace{0.5cm}

Now we start to consider the gravitational perturbation around the
tidal charged black hole. The first order perturbation of the
Eq.(\ref{e2}) should be $\delta R_{\mu\nu}=-\delta E_{\mu\nu}$. For
the maximally symmetric black holes in high dimensions, a master
equation can be obtained for gravitational perturbations
\cite{Kodama}. However due to the lack of the knowledge of the exact
5D bulk metric in our case, we cannot determine the $\delta
E_{\mu\nu}$ expression. In \cite{s11} \cite{abdalla}, a simple
assumption $\delta E_{\mu\nu}=0$ was adopted and justified in a
region where the perturbation energy does not exceed the threshold
of the Kaluza-Klein massive modes \cite{abdalla2}. On the other
hand, the similarity to the metric of the RN black hole sheds
another light on the problem. Based on the fact that the
characteristic of $E_{\mu\nu}$ on the brane is almost the same as
the form of the energy -momentum tensor of electromagnetic field
,{\it i.e.} its symmetry and tracelessness, we can apply the
gravitational perturbation theory of the RN black hole to the
braneworld black hole with tidal charge. Classifying the
perturbation into axial and polar components and separating the
angular dependance as done in \cite{mellor&moss}, we can obtain the
axial perturbation which reads
\begin{equation} \label{e27}
\Lambda ^2 Z_i^ -   = V_i^ -  Z_i^ -, \quad (i=1,2)
\end{equation}
where
\begin{equation} \label{e28}
\Lambda ^2  = \frac{{\partial ^2 }}{{\partial r_*^2 }} -
\frac{{\partial ^2 }}{{\partial t^2 }}.
\end{equation}
The effective potential reads:
\begin{equation} \label{e31}
V_i^ -   = \frac{\Delta }{{r^5 }}\left[ {2(n + 1)r - p_j (1 +
\frac{{p_i }}{{2nr}})} \right]\quad (i,j = 1,2;i \ne j)
\end{equation}
where
\begin{equation} \label{e29}
n = \frac{1}{2}(l - 1)(l + 2)
\end{equation}

\begin{equation} \label{e30}
\begin{array}{l}
 p_1  = 3M + \sqrt {9M^2  - 8nQ}  \\
 \\
 p_2  = 3M - \sqrt {9M^2  - 8nQ}  \\
 \end{array}
\end{equation}
and $ \Delta  = r^2  - 2Mr - Q$.

The polar perturbations are described as
\begin{equation} \label{e32}
\Lambda ^2 Z_i^ +   = V_i^ +  Z_i^ +  \quad (i = 1,2)
\end{equation}
and the effective polar potentials are
\begin{equation} \label{e33}
\begin{array}{l}
 V_1^ +   = \frac{\Delta }{{r^5 }}\left[ {U + \frac{1}{2}(p_1  - p_2 )W} \right] \\
 \\
 V_2^ +   = \frac{\Delta }{{r^5 }}\left[ {U - \frac{1}{2}(p_1  - p_2 )W} \right]. \\
 \end{array}
\end{equation}
\begin{figure}[ht]
\vspace*{0cm}
\begin{minipage}{0.4\textwidth}
\resizebox{0.9\linewidth}{!}{\includegraphics*{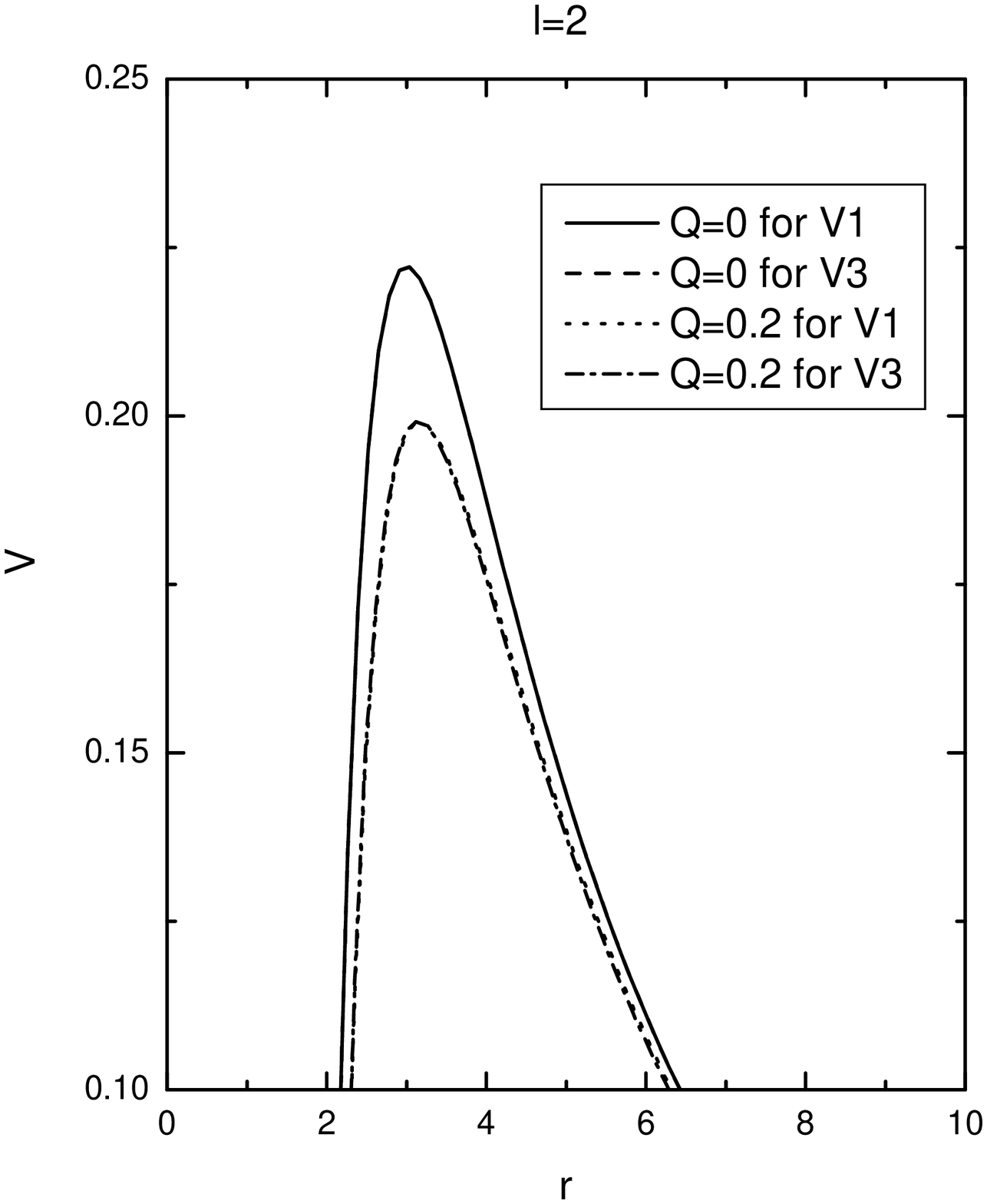}}
\nonumber
\end{minipage}\nonumber
\begin{minipage}{0.4\textwidth}
\vspace*{0.0cm}
\resizebox{0.9\linewidth}{!}{\includegraphics*{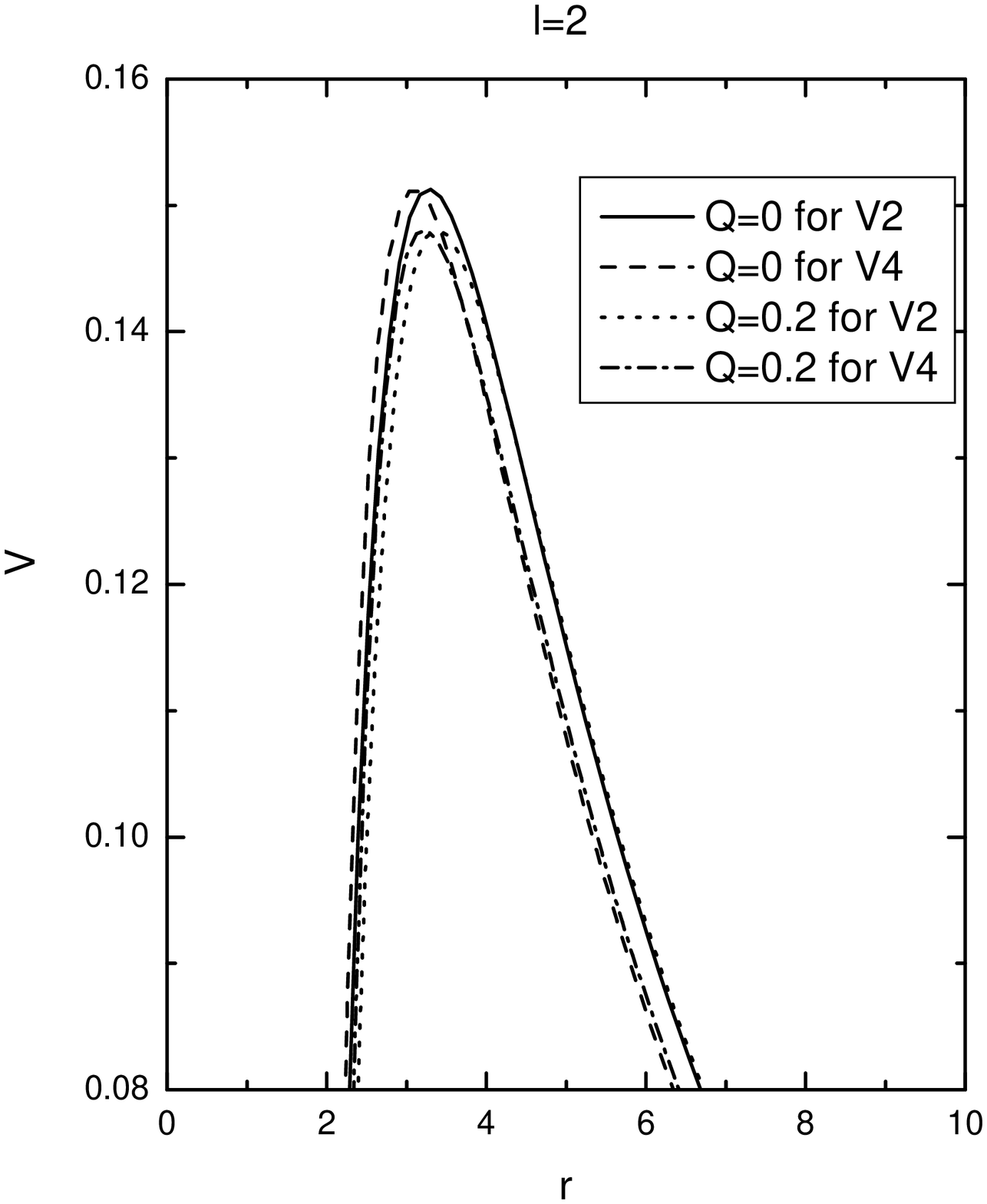}}
\nonumber
\end{minipage} \nonumber
\begin{minipage}{0.4\textwidth}
\vspace*{0.0cm}
\resizebox{0.9\linewidth}{!}{\includegraphics*{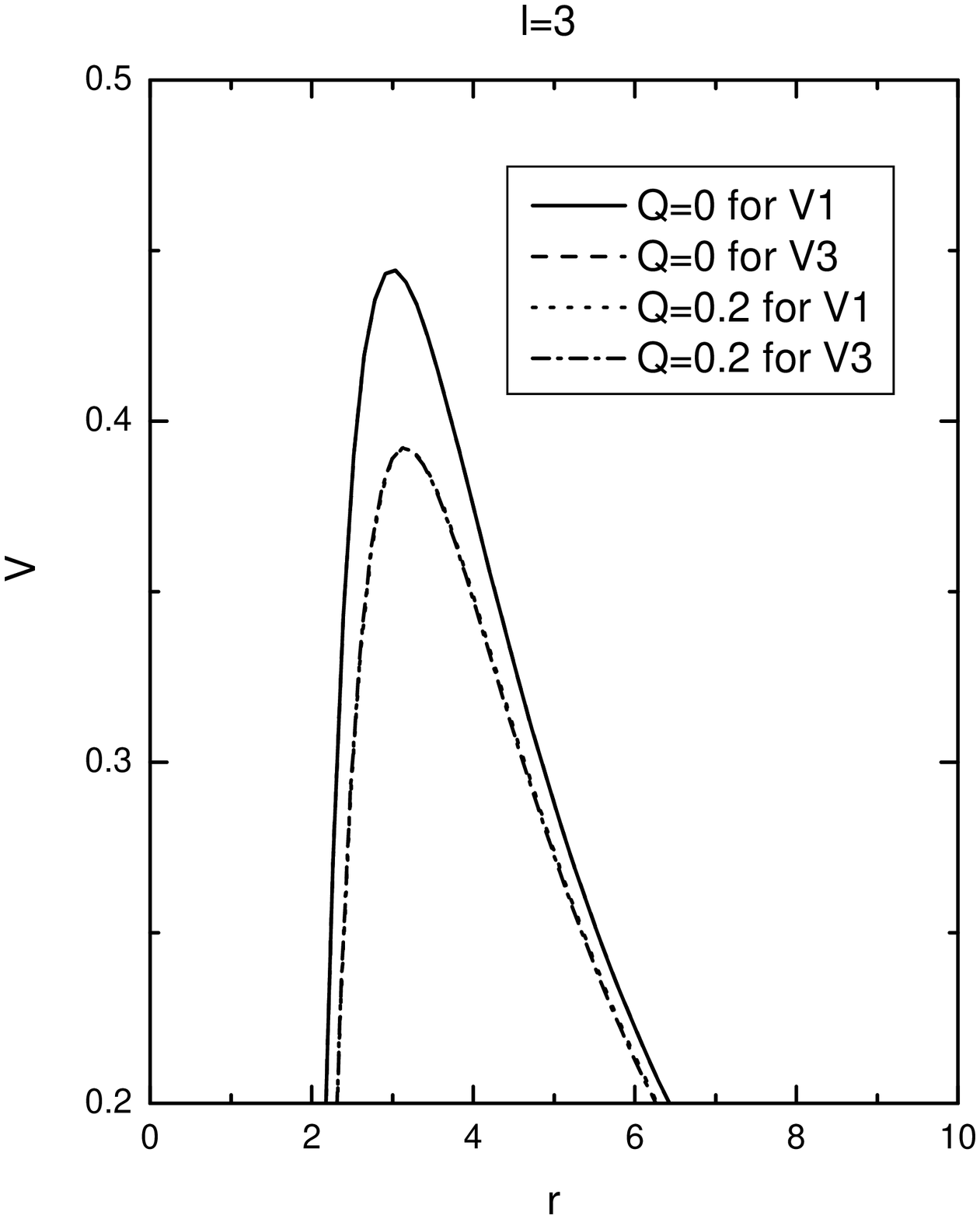}}
\nonumber
\end{minipage} \nonumber
\begin{minipage}{0.4\textwidth}
\vspace*{0.0cm}
\resizebox{0.9\linewidth}{!}{\includegraphics*{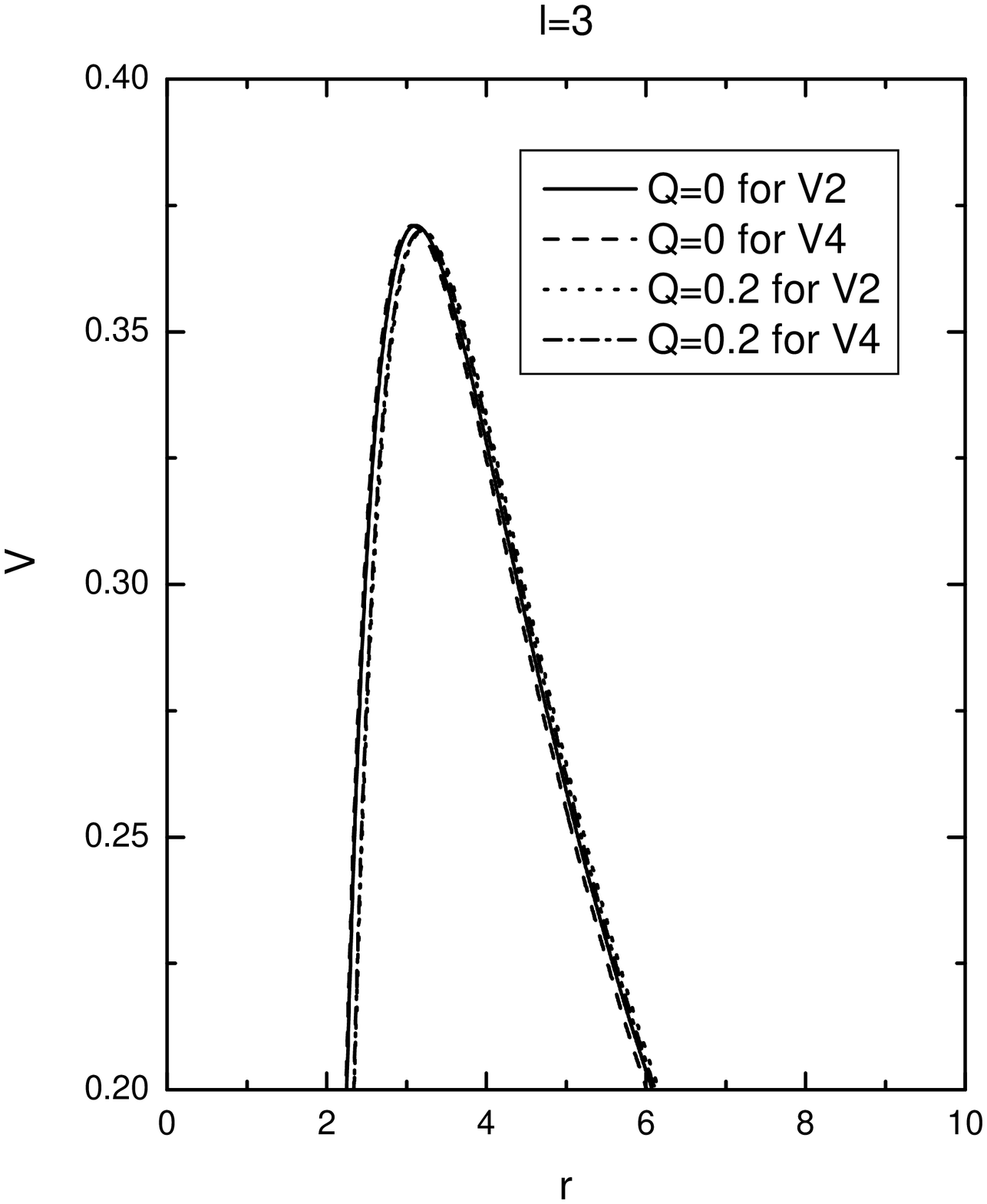}}
\nonumber
\end{minipage} \nonumber
\caption{{The comparison of the effective potential of the
gravitational perturbation between the Schwarzschild black hole and
the brane world black hole. We have taken black hole mass
$M=1$.}}\label{f5}
\end{figure}

Here $U$ and $W$ are given by
\begin{equation} \label{e34}
\begin{array}{l}
 W = \frac{\Delta }{{r\varpi ^2 }}(2n + 3M) + \frac{1}{\varpi }(nr + M) \\
 \\
 U = (2n + 3M)W + (\varpi  - nr - M) - \frac{{2n\Delta }}{\varpi } \\
 \end{array}
\end{equation}
with $ \varpi  = nr + 3M + \frac{{2Q}}{r}$. It is noticed that when
$ Q > \frac{{9M^2 }}{{8n}}\quad (l\ne 0)$, $p_1$ and $p_2$ will be
complex and the appearance of the imaginary part in the effective
potentials will cause a series of interesting physical consequences.
When $M=1$, the upper limits $Q_{up}$ are $0.5625$ for $l=2$ and
$0.225$ for $l=3$. We will discuss it briefly in the next section
and here we just consider the case when $ Q \le \frac{{9M^2
}}{{8n}}\quad (l\ne 0)$. Eqs.(\ref{e27}) and (\ref{e32}) can be
written as
\begin{equation} \label{e35}
\frac{{\partial ^2 Z_i }}{{\partial r_*^2 }} - \frac{{\partial ^2
Z_i }}{{\partial t^2 }} = V_i Z_i  \quad (i=1,2,3,4)
\end{equation}
by using Eq.(\ref{e9}) and the separation of time dependence
$e^{-ikt}$. The potentials $V_i$ are $V_1^-$, $V_2^-$, $V_1^+$ and
$V_2^+$ when $i$ runs from $1$ to $4$. And $Z_i$ are defined
similarly. Although the mathematical forms of $V_i$ are very
different, the numerical calculation shows that the difference
between the value $V_1$ and $V_3$ is extremely small and the same
situation happens for $V_2$ and $V_4$ as shown in Fig.\ref{f5}.
Actually, if we take $Q \to -Q^2$ in the Eq.(\ref{e5}), $V_1$(or
$V_3$) represents the potential of the pure electromagnetic
perturbation and $V_2$(or $V_4$) of gravitational perturbation in
the RN black hole. In the stage of the brane-world black hole with a
small $Q$, the role played by the static electrical charge as the
pure electromagnetic perturbation in RN black hole is now acted by
the tidal charge as the perturbation relevant tightly to that
induced by the bulk effects, with the consideration that
the exact bulk metric is still unknown and its effects are
parameterized by $Q$ on the brane. Therefore it is believed that
$V_2$(or $V_4$) stands for the gravitational perturbation mainly
caused by the fluctuations of the brane, while $V_1$(or $V_3$) for
the potential contributed by the perturbation dominated by the
fluctuations of the bulk effects projected on the brane.

Following the same steps as in the case of the scalar field
perturbation, we can compute the QNMs of the gravitational
perturbations. Due to the similarity between the potentials $V_1$,
$V_3$ and $V_2$, $V_4$, their QNM results are extremely the same.
For the negative tidal charged black hole, the quasinormal ringing
behavior is separated from the positive charged black hole by the
border when $Q=0$, which is the 4D Schwarzschild black hole cases.
When the tidal charge becomes more negative, we observed the same
behavior as that of the scalar perturbation, the gravitational
perturbation will have less oscillation and last longer. This
behavior is monotone with the increase of $|Q|$ when the tidal
charge is negative. For the positive tidal charge, we observed the
same QNM behavior as that of the RN black hole disclosed in
\cite{berti}.
\vspace{0.5cm}
\begin{center}

$ \omega= \omega_R + i \omega_I$

\begin{tabular}{l|l|l|l|l|l|r} \hline
 & $l=2$ & $l=3$ \\ \hline \hline
RN. $Q=-0.5$ for $V_1$,$V_3$ &   $0.54 - i 0.086$   &  $0.76 - i 0.086$   \\
Sch. $Q=0$   for $V_1$,$V_3$  &   $0.46 - i 0.083$   &  $0.66 - i 0.083$   \\
Tdl. $Q=0.2$ for $V_1$,$V_3$ &   $0.43 - i 0.081$   &  $0.62 - i 0.081$   \\
\hline\hline
RN. $Q=-0.5$ for $V_2$,$V_4$ &   $0.39 - i 0.078$   &  $0.63 - i 0.081$   \\
Sch. $Q=0$   for $V_2$,$V_4$   &   $0.37 - i 0.077$   &  $0.60 - i 0.080$   \\
Tdl. $Q=0.2$ for $V_2$,$V_4$ &   $0.37 - i 0.077$   &  $0.60 - i 0.080$   \\
\hline
\end{tabular}

\vspace{0.1cm} \large{Table II}

\vspace{0.1cm} {\it Quasinormal frequencies of the gravitational
perturbation around the Schwarzchild and tidal charge black
holes.}
\end{center}
\vspace{0.5cm}

\section{Absorption and Emission Spectra}
In this section, we are going to discuss the emission of
brane-localized scalars and gravitons. We will present the numerical
results and comment on the extra dimensional influence on the
particles emitted on the brane.

\subsection{The massless scalar field}
Taking $ \Psi _l  = rR_l (r)e^{ - ikt}$ and substituting it into
Eq.(\ref{e8}), we can write the radial part of KG equation as
\begin{equation} \label{e10}
(r - r_ +  )^2 (r - r_ -  )^2 R_l '' + [2r - (r_ +   + r_ -  )](r -
r_ +  )(r - r_ -  )R_l ' + \left\{ {r^4 k^2  - l(l + 1)(r - r_ + )(r
- r_ -  )} \right\}R_l = 0.
\end{equation}
Following the method of Persides\cite{persides}, the analytic
series solution of Eq.(\ref{e10}) near the horizon can be
expressed as
\begin{equation} \label{e11}
R_l(r)= (r - r_ +  )^{ \rho  } \sum\limits_{n = 0}^\infty {g_{l,n}
(r - r_ +  )^n } \quad  r \to r_+,
\end{equation}
where the series of constants $g_{l,n}$ is determined by the
iteration process substituting Eq.(\ref{e11}) into the radial
equation Eq.(\ref{e10}). The index $\rho$ is given by the index
equation according to the Frobenius-Fuchs theorem in Mathematics
\begin{equation} \label{e12}
\rho ^2  + \frac{{k^2 r_ + ^4 }}{{(r_ +   - r_ -  )^2 }} = 0.
\end{equation}
In fact, the index equation depends only on the metric of the
black hole and has nothing to do with the types of fields and
their effective potentials. Hence, we take
\begin{equation} \label{e13}
\rho= - i\kappa _ +   = -i \frac{{kr_ + ^2 }}{{r_ +   - r_ -  }}
\end{equation}
and the negative sign indicates that there is only ingoing wave near
the horizon. In the asymptotically flat region, the solution of
Eq.(\ref{e10}) consists of the outgoing and ingoing waves
\begin{equation} \label{e14}
R_l (r) = (l+1/2)[f_l^{( - )} F_{l( + )}  + f_l^{( + )} F_{l( - )}],
\end{equation}
where $f_l^{(-)}$ and $f_l^{(+)}$ are the coefficients. $F_{l
(+)}$ and $F_{l(-)}$ correspond to the ingoing and outgoing waves
respectively and have the form
\begin{equation} \label{e15}
F_{l( \pm )} (r) = ( \pm i)^{l + 1} \exp \left[ { \mp ikr_* }
\right]\sum\limits_{n = 0}^\infty  {\tau _{n(\pm)}  (r - r_ +
)^{-(n+1)} } \quad r \to \infty
\end{equation}
with $\tau_{0(\pm)} =1$. Although the total series of the
constants $g_{l,n}$ and $\tau_{n(\pm)}$ can be obtained in the
iteration relations, the dominant terms are those of $n=0$ in
Eq.(\ref{e11})and Eq.(\ref{e15}). The solutions of wave are
rewritten as
\begin{equation}\label{e16}
\begin{array}{l}
 R_l  \approx g_{l,0} (r - r_ +  )^{ - i\kappa _ +  } [1 + O(r - r_ +  )]\quad r \to r_ +   \\
\\
 R_l  \approx i^{l + 1} f_l^{(-)} \frac{{2l + 1}}{{2r}}\left[ {e^{ - ikr_* }  - ( - 1)^l S_l e^{ + ikr_* } } \right] + O(\frac{1}{{r^2 }})\quad r \to \infty,  \\
 \end{array}
\end{equation}
where $ S_l  = \frac{{f_l ^{( + )} }}{{f_l ^{( - )} }} $ is the
partial scattering amplitude. Introducing a phase shift $\delta_l$
in $S_l=e^{2 i \delta_l}$ \cite{park}\cite{s11}, the wave solution
in the asymptotic region is
\begin{equation} \label{e17}
R_l  \approx f_l^{(-)} \frac{{2l + 1}}{r}e^{i\delta _l } Sin\left[
{kr_* - \frac{\pi }{2}l + \delta _l } \right] + O(\frac{1}{{r^2
}})\quad r \to \infty.
\end{equation}

In the spirit of scattering theory of quantum mechanics (Q.M.), the
ingoing wave coming from the infinity partly penetrates the
potential barrier and is absorbed by the black hole, while the other
part is scattered by the barrier and return to the infinity.
Therefore, the compatibility among the coefficients $g_{l,0}$,
$f_l^{-}$ and the scattering amplitude $S_l$ must be found for the
true solution of radial equation (\ref{e10}). Then the absorption
cross section and the emission rate can be calculated. For this
reason, it is necessary to discuss the Wronskians of Eq.(\ref{e10}).
As done in the Q.M., it is easy to get the equation of Wronskians
\begin{equation} \label{e19}
W' + \frac{{2r - r_ +   - r_ -  }}{{(r - r_ +  )(r - r_ -  )}}W = 0
\end{equation}
from Eq.(\ref{e10}) with the definition
\begin{equation} \label{e18}
W[R_l ^* ,R_l ] = R_l ^* \frac{{\partial R_l }}{{\partial r}} - R_l
\frac{{\partial R_l ^* }}{{\partial r}}.
\end{equation}
The analytic solution of Eq.(\ref{e19}) is
\begin{equation} \label{e20}
W = \frac{C}{{(r - r_ +  )(r - r_ -  )}}.
\end{equation}
Since the 'current' density is proportional to the Wronskian
\cite{sanchez}, the constant $C$ plays a role of the 'flux'
conversation analogous to what happens in Q.M. and the additional
denominator $(r - r_ + )(r - r_ - )$ shows the effect of the curved
spacetime background. Substituting the solution near the horizon
Eq.(\ref{e16}) and that in the asymptotic region Eq.(\ref{e17}) into
Eq.(\ref{e18}) and Eq.(\ref{e20}), we have
\begin{equation} \label{e21}
C =  - 2ik\left| {g_{l,0} } \right|^2 r_ +  ^2  =  - ik\left|
{f_l^{( - )} } \right|^2 (2l + 1)^2 e^{ - 2\beta _l } \rm
Sinh(2\beta _l ),
\end{equation}
and
\begin{equation} \label{e22}
\left| {g_{l,0} } \right|^2  = \left| {f_l^{( - )} } \right|^2
\frac{{(l + 1/2)^2 }}{{r_ +  ^2 }}(1 - e^{ - 4\beta _l } ),
\end{equation}
where $\beta_l = \rm Im[\delta_l]$. The partial absorption cross
section $\sigma_l$ is defined in the terms of $g_{l,0}$
\cite{sanchez} with unit amplitude $|f_l^{(-)}|^2=1$ of the ingoing
wave as
\begin{equation} \label{e23}
\sigma _l  = 4\pi r_ + ^2 \frac{{\left| {g_{l,0} } \right|^2 }}{{(2l
+ 1)k^2 }} = \frac{\pi }{{k^2 }}(2l + 1)(1 - e^{ - 4\beta _l } ).
\end{equation}
The partial emission spectrum is given by the Hawking formula
\begin{equation} \label{e24}
\Gamma _l  = \frac{{dH_l }}{{dk}} = \frac{{k^3 \sigma _l }}{{2\pi ^2
(e^{k/T}  - 1)}}
\end{equation}
at the temperature of the black hole $ T = \frac{{r_ +   - r_ -
}}{{4\pi r_ + ^2 }}$. The total absorption cross section and
emission spectrum are $ \sigma _{tot}  = \sum\limits_l {\sigma _l }$
and $ \Gamma _{tot}  = \sum\limits_l {\Gamma _l }$ respectively.

\begin{figure}[ht]
\vspace*{0cm}
\begin{minipage}{0.3\textwidth}
\resizebox{1.1\linewidth}{!}{\includegraphics*{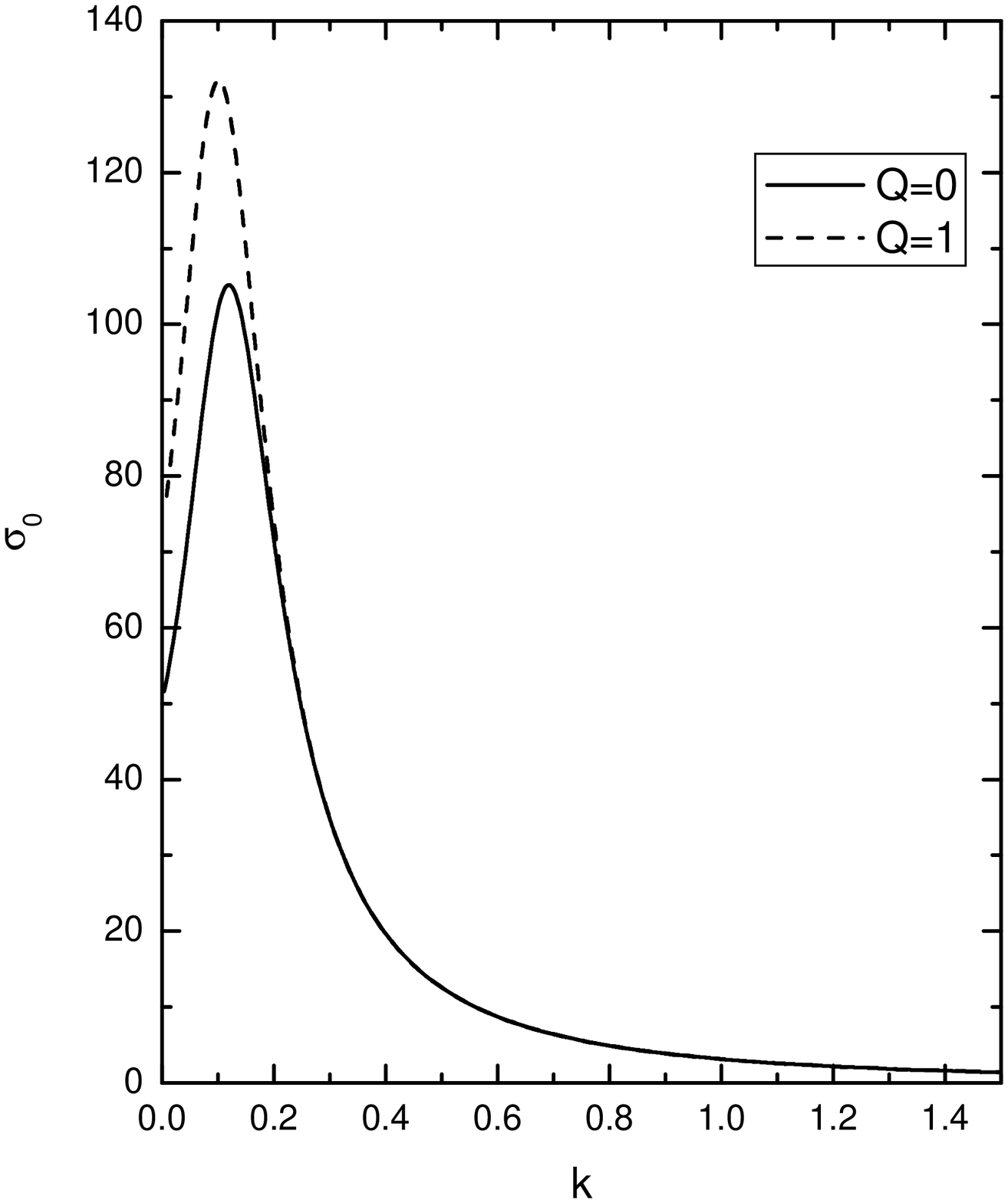}}
\nonumber
\end{minipage}\nonumber
\begin{minipage}{0.3\textwidth}
\vspace*{0.0cm}
\resizebox{1.1\linewidth}{!}{\includegraphics*{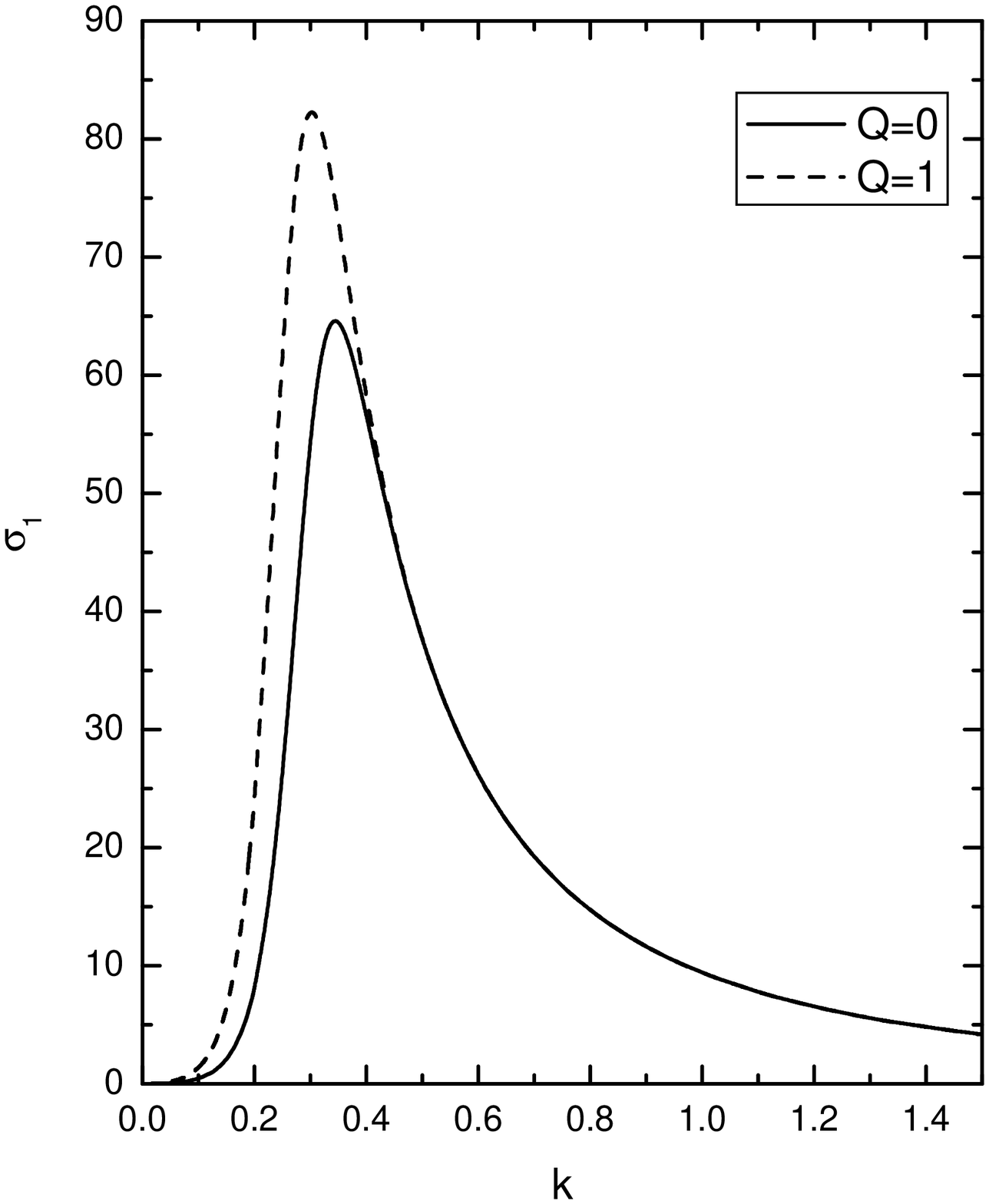}}
\nonumber
\end{minipage} \nonumber
\begin{minipage}{0.3\textwidth}
\vspace*{0.0cm}
\resizebox{1.1\linewidth}{!}{\includegraphics*{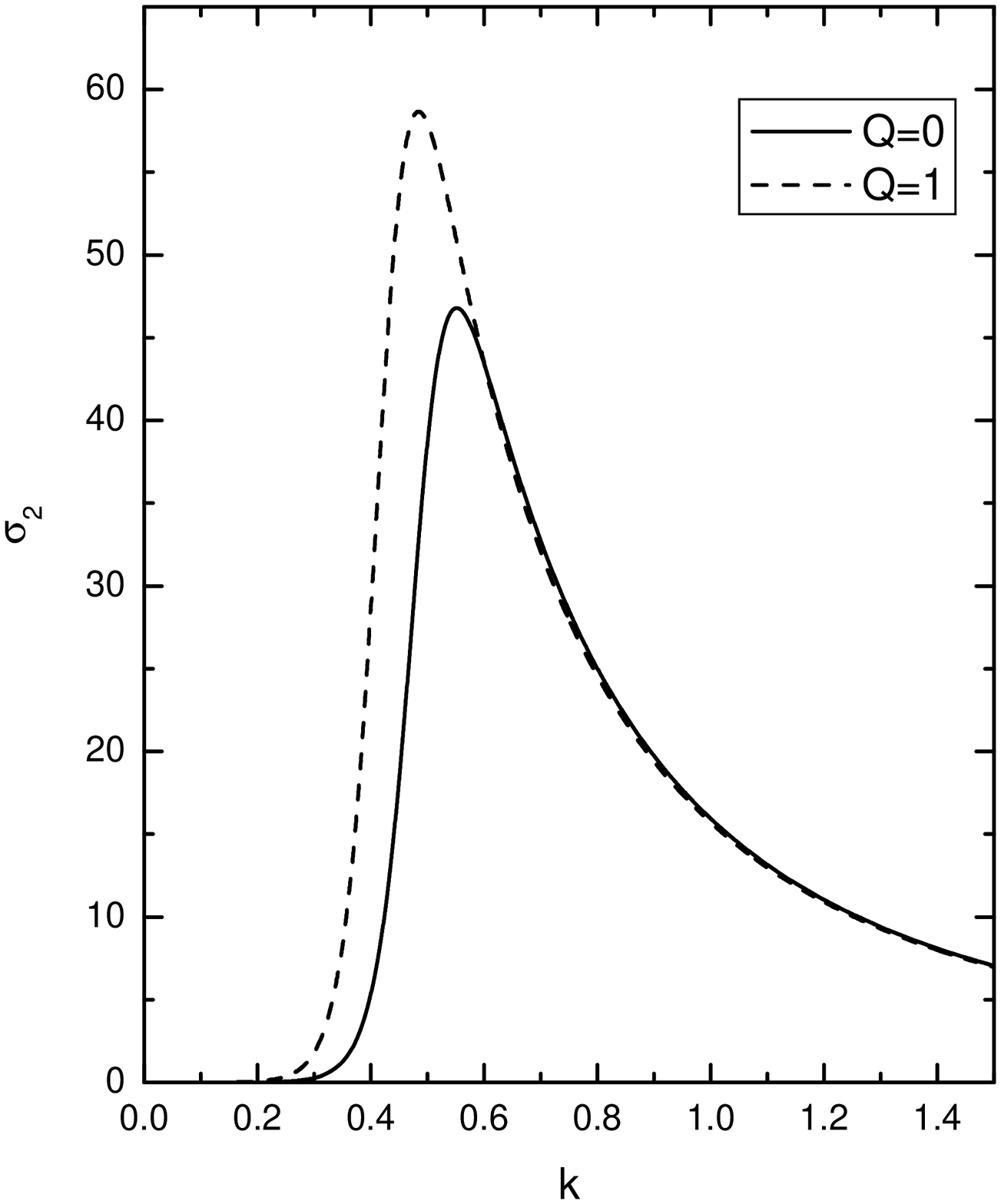}}
\nonumber
\end{minipage} \nonumber
\caption{{The spectrum of the partial absorption cross sections
$\sigma_l$ for different angular indexes in the Schwartzchild black
hole and the brane-world black hole. From the left to the right we
choose $l=0$, $l=1$ and $l=2$ respectively.}}\label{f2}
\end{figure}

Now we introduce the numerical method to obtain the imaginary part
$\beta_l$ in Eq.(\ref{e23}) of the phase shift, which is the key to
the question of the absorption cross section and emission spectrum.
This computation has also been done in \cite{park}\cite{s11}. The
Wronskians of the ingoing and outgoing wave in the asymptotic region
have the following property
\begin{equation} \label{e25}
\begin{array}{l}
 W[F_{l( + )} ,R_l ] = F_{l( + )} R'_l  - R_l F'_{l( + )}  = (l+1/2) f_l^{( + )} W[F_{l( + )} ,F_{l( - )} ] \\
 \\
 W[F_{l( - )} ,R_l ] = F_{l( - )} R'_l  - R_l F'_{l( - )}  =  - (l+1/2) f_l^{( - )} W[F_{l( + )} ,F_{l( - )} ], \\
 \end{array}
\end{equation}
where $ W[F_{l( + )} ,F_{l( - )} ] = F_{l( + )} F'_{l( - )}  - F_{l(
- )} F'_{l( + )}$. The imaginary part $\beta_l$ can be expressed as
\begin{equation} \label{e26}
\beta _l  =  - \frac{1}{2}{\mathop{\rm Im}\nolimits} \left[ {i\ln
S_l } \right] =  - \frac{1}{2}{\mathop{\rm Im}\nolimits} \left[
{i\ln \frac{{f_l^{( + )} }}{{f_l^{( - )} }}} \right] =  -
\frac{1}{2}{\mathop{\rm Im}\nolimits} \left[ {i\ln \left( { -
\frac{{W[F_{l( + )} ,R_l ]}}{{W[F_{l( - )} ,R_l ]}}} \right)}
\right].
\end{equation}
In practice, we take $g_{l,0}=1$ for simplicity by considering that
the partial scattering amplitude $S_l$ is only determined by the
ratio of $f_l^{(+)}$ and $f_l^{(-)}$ and the real part of $\delta_l$
is not important in this problem. Using the first equation in
(\ref{e16}) as a boundary condition, one can integrate the radial
equation Eq.(\ref{e10}), calculate numerically the ratio of the two
Wronskians from Eq.(\ref{e25}) with the second equation of
(\ref{e16}) in the asymptotic region and finally get the value of
$\beta_l$ from the (\ref{e26}).

\begin{figure}
\resizebox{0.5\linewidth}{!}{\includegraphics*{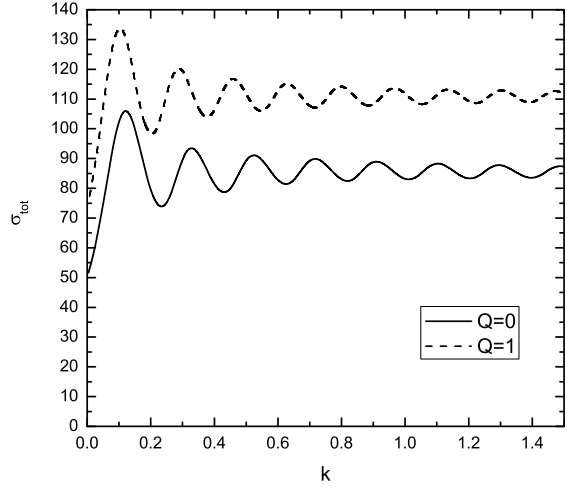}}
\caption{The spectrum of the total absorption cross section
$\sigma_{tot}$ in the Schwartzchild black hole and the brane-world
black hole.}
\label{f3}
\end{figure}

We have calculated the spectrum of absorption cross section in black
holes with the mass $M=1$ in both Schwarzschild and brane-world
cases with the tidal charge. Fig.\ref{f2} shows the spectrum of the
partial absorption cross sections $\sigma_l$ at different $l$. We
can see that $\sigma_0$ equals to the surface area of the black hole
$4 \pi r_+^2$ as $k \to 0$, which is a universal property in
low-energy region for the S-wave. All partial cross sections reach
their peaks first and then decrease in the high-energy region. It is
noticed that the $\sigma_l$ on the brane-world black hole is always
bigger than that in the Schwarzschild black hole because of its
lower potential barriers with negative tidal charge as shown in
Fig.\ref{f1}. This indicates that the scalar field around the
negative tidal charge brane-world black hole is absorbed by the
black hole more easily due to the decrease of the potential barrier.
The fact that more absorption is found around the black hole with
negative tidal charge supports the argument that the bulk effect
strengthen the gravitational field \cite{Tdl}. In Fig.\ref{f3}, we
show the spectrum of the total absorption cross section
$\sigma_{tot}$ which has the limiting value $27 \pi r_+^2 /4$
\cite{s11} as $k \to \infty$.

\begin{figure}
\resizebox{0.5\linewidth}{!}{\includegraphics*{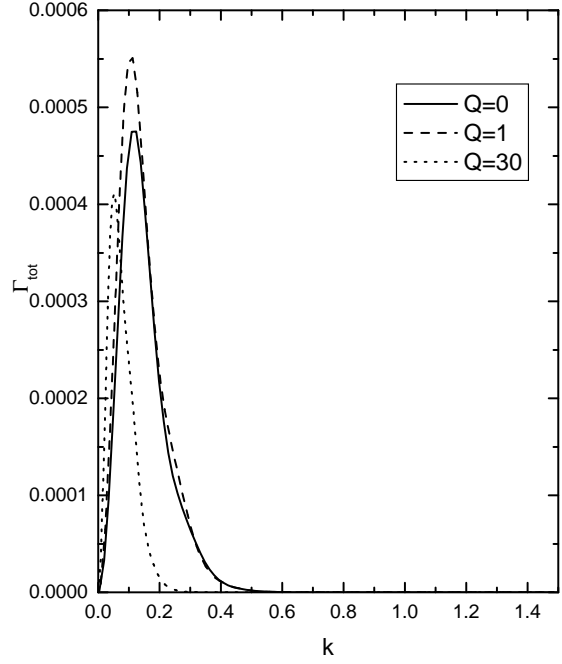}}
\caption{The total emission spectrum $\Gamma_{tot}$ in the
Schwartzchild black hole and the brane-world black hole.}
\label{f4}
\end{figure}

According to Eq.(\ref{e24}), we have done the calculation of the
total emission spectrum as shown in Fig.\ref{f4}. Unlike the
oscillation pattern in the total absorption cross section in
Fig.\ref{f3}, the spectrum of emission reaches its peak and falls
monotonely. The high $l$ modes of $\sigma_l$ seems suppressed in the
Hawking radiation, since the peak of $\Gamma_{tot}$ for high $l$ is
located at almost the same position of the peak of $\sigma_0$. The
negative tidal charge $Q$ enlarges the black hole horizon and
decreases the temperature of the brane world black hole compared to
that of the 4D Schwarzschild black hole. However, the total emission
spectrum does not decrease simply following the temperature as shown
in the Fig.\ref{f4}. There is a critical value $Q_h \approx 20$ for
the black hole mass $M=1$. When $Q>Q_h$ with the much lower
temperature, the maximum value of the emission spectrum returns
below that of Schwarzschild black hole. But when $0 < Q < Q_h$, the
emission spectrum is enhanced with the increase of $|Q|$ and could
be bigger than that of Schwarzschild black hole with the same mass.
This behavior of the emission spectrum due to different values of
negative tidal charge is the result of the competition between the
enhancement of the absorption cross section in the numerator and the
decreasing temperature in the denominator in (\ref{e24}). It changes
the usual picture of the Hawking radiation that lower temperature
leads to lower emission and shows the complexity of the influences
of the bulk effects on the Hawking radiation.

\subsection{The brane-localized gravitational perturbation}
We would like to extend our discussion on the absorption cross
section and emission spectrum to the gravitational perturbation in
the brane-world black hole background. We will concentrate on the
brane-localized perturbation, where the flux in the brane
Eq.(\ref{e37}) is conserved. As what happens in the scalar field,
when the tidal charge becomes more negative, the heights of the
potential barriers for $V_1$ (or $V_3$) and $V_2$ (or $V_4$) are
suppressed. The decrease of the potential of $V_2$ (or $V_4$) is not
as much as that of $V_1$ (or $V_3$). It is easier to influence the
potential $V_1$ (or $V_3$) than $V_2$ (or $V_4$) by the bulk effects
because of their different origins.

\begin{figure}[ht]
\vspace*{0cm}
\begin{minipage}{0.4\textwidth}
\resizebox{0.9\linewidth}{!}{\includegraphics*{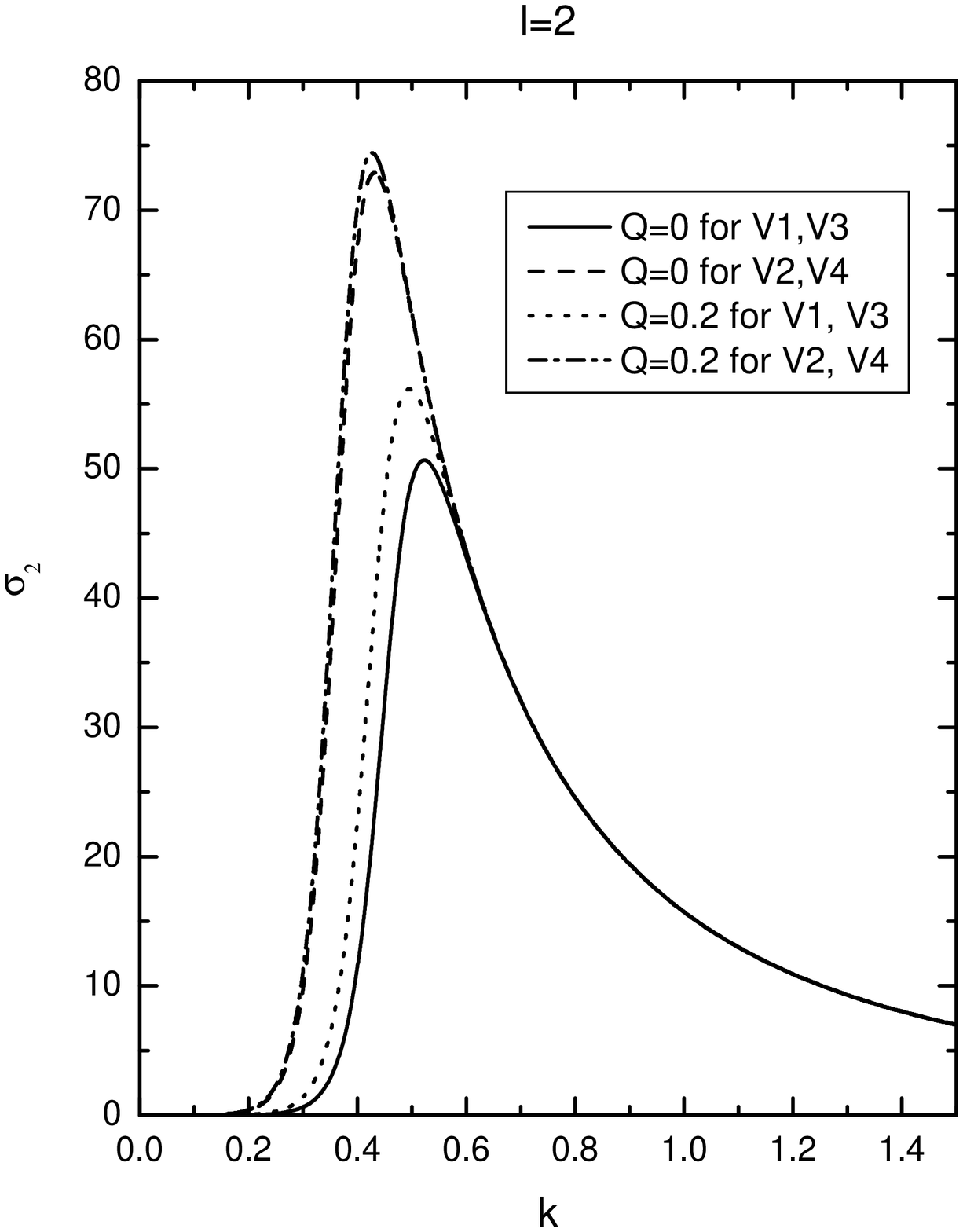}}
\nonumber
\end{minipage}\nonumber
\begin{minipage}{0.4\textwidth}
\vspace*{0.0cm}
\resizebox{0.9\linewidth}{!}{\includegraphics*{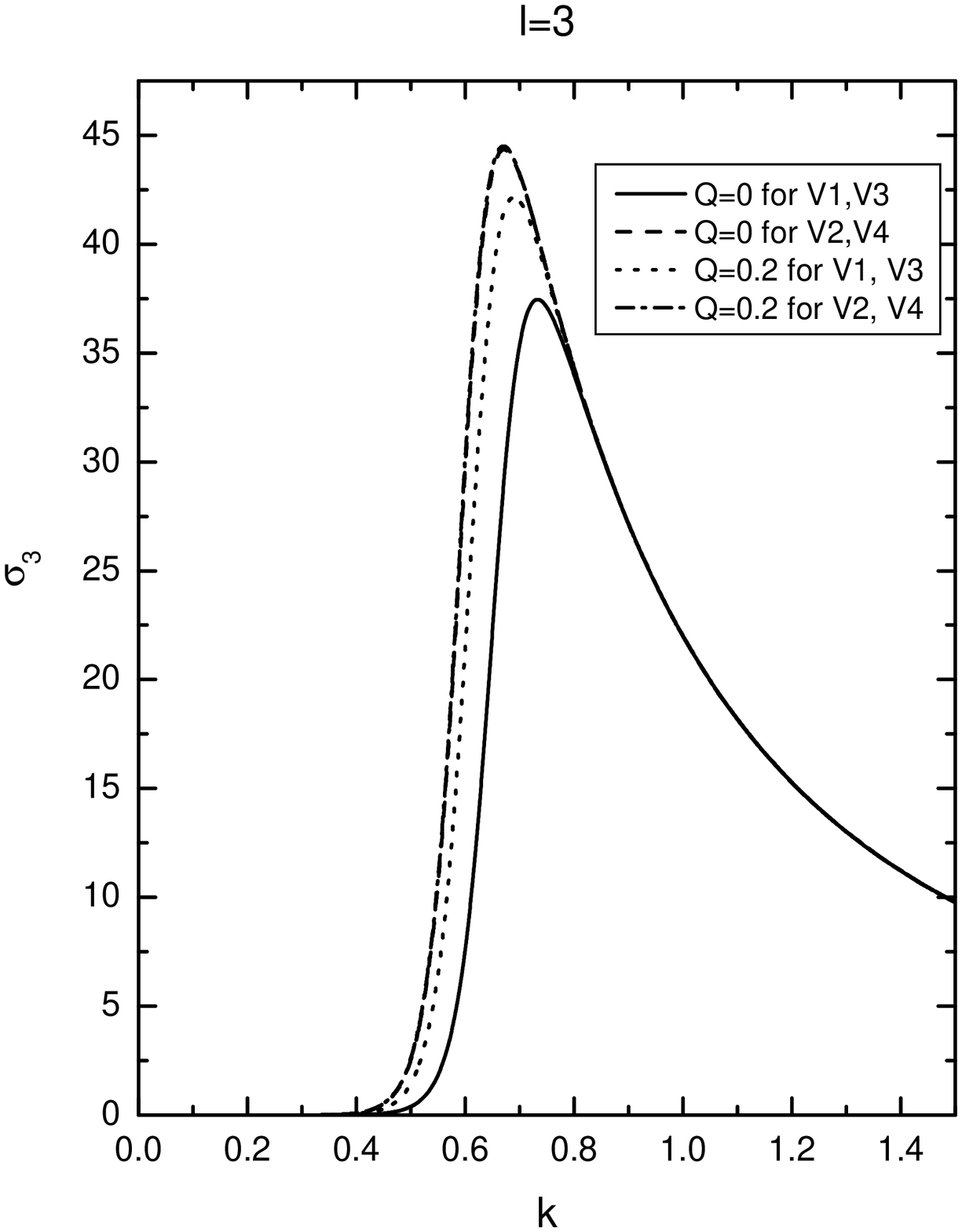}}
\nonumber
\end{minipage} \nonumber
\caption{{The spectrum of the partial absorption cross section
$\sigma_l$ with $M=1$ by taking $l=2$ and $l=3$ in the Schwartzchild
black hole and the brane-world black hole.}}\label{f6}
\end{figure}

By recovering the normal coordinates and separating out the time
dependence, the perturbation equation (\ref{e35}) can be expressed
as
\begin{equation} \label{e44}
f^2 Z_i '' + ff'Z_i ' + k^2 Z_i  = V_i Z_i  \quad (i=1,2,3,4).
\end{equation}
Following the steps in the last subsection, we can get the same
index $\rho$ as in the Eq.(\ref{e13}) near the black hole horizon
and in the asymptotic region $\rho=0$
\begin{equation}\label{e36}
\begin{array}{l}
 Z_i  \approx g_{l,0}r_+ (r - r_ +  )^{ - i\kappa _ +  } [1 + O(r - r_ +  )]\quad r \to r_ +   \\
\\
 Z_i  \approx f_l^{(-)} (2l + 1)e^{i\delta _l } Sin\left[
{kr_* - \frac{\pi }{2}l + \delta _l } \right] + O(\frac{1}{{r
}})\quad r \to \infty. \\
 \end{array}
\end{equation}
The Wronskian of Eq.(\ref{e35}) as the definition of Eq.(\ref{e19})
is
\begin{equation} \label{e37}
f^2 W' + ff'W = 0 \Rightarrow W = \frac{C}{f} = \frac{{r^2 C}}{{(r -
r_ +  )(r - r_ -  )}}
\end{equation}
by combining the metric Eq.(\ref{e5}). $C$ is still the constant
with the meaning of the 'flux' conservation of the gravitational
perturbation in the brane-world black hole and its value is
\begin{equation} \label{e38}
C =  - 2ik\left| {g_{l,0} } \right|^2 r_ +  ^2  =  - ik\left|
{f_l^{( - )} } \right|^2 (2l + 1)e^{ - 2\beta _l } \rm Sinh(2\beta
_l )
\end{equation}
according to Eq.(\ref{e36}). Expressions of the partial absorption
cross section Eq.(\ref{e23}) and the emission spectrum
Eq.(\ref{e24}) are still valid in the gravitational perturbation.
Using the numerical method introduced above, we have calculated
$\sigma_l$ and the results are shown in Fig.\ref{f6}. The absorption
cross sections for $V_1$ and $V_3$ (or $V_2$ and $V_4$) overlap each
other because of the identical numerical behaviors of the
potentials. The suppression of the potential $V_i$ barriers makes
the absorption cross section larger. The influences of the tidal
charge are much bigger for $V_1$(or $V_3$) than that for $V_2$(or
$V_4$) due to their different perturbative origins as discussed in
the last section. Comparing to the 4D Schwarzschild black hole, the
contrasts of the increase of the peak value due to the negative
tidal charge with the bulk influence $|\sigma_l^{Tdl}-\sigma_l^{Sch}
| / \sigma_l^{Sch}$ are $11 \%$ and $2\%$ for $V_1$(or $V_3$) and
$V_2$(or $V_4$) respectively when $l=2$. At $l=3$ the contrasts are
$13 \%$ and $0.4 \%$ for $V_1$($or V_3$) and $V_2$(or $V_4$)
respectively.

\begin{figure}[ht]
\vspace*{0cm}
\begin{minipage}{0.4\textwidth}
\resizebox{0.9\linewidth}{!}{\includegraphics*{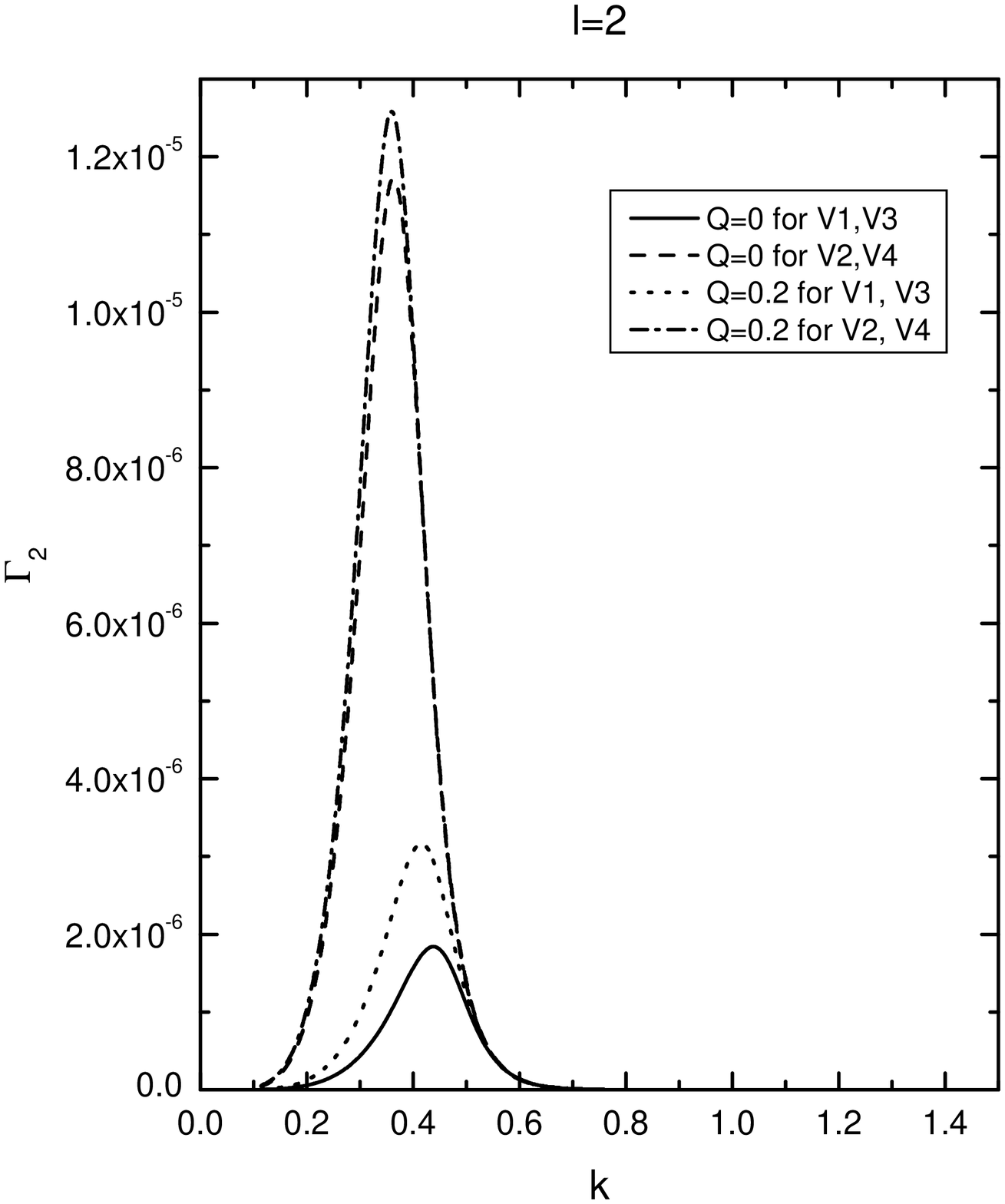}}
\nonumber
\end{minipage}\nonumber
\begin{minipage}{0.4\textwidth}
\vspace*{0.0cm}
\resizebox{0.9\linewidth}{!}{\includegraphics*{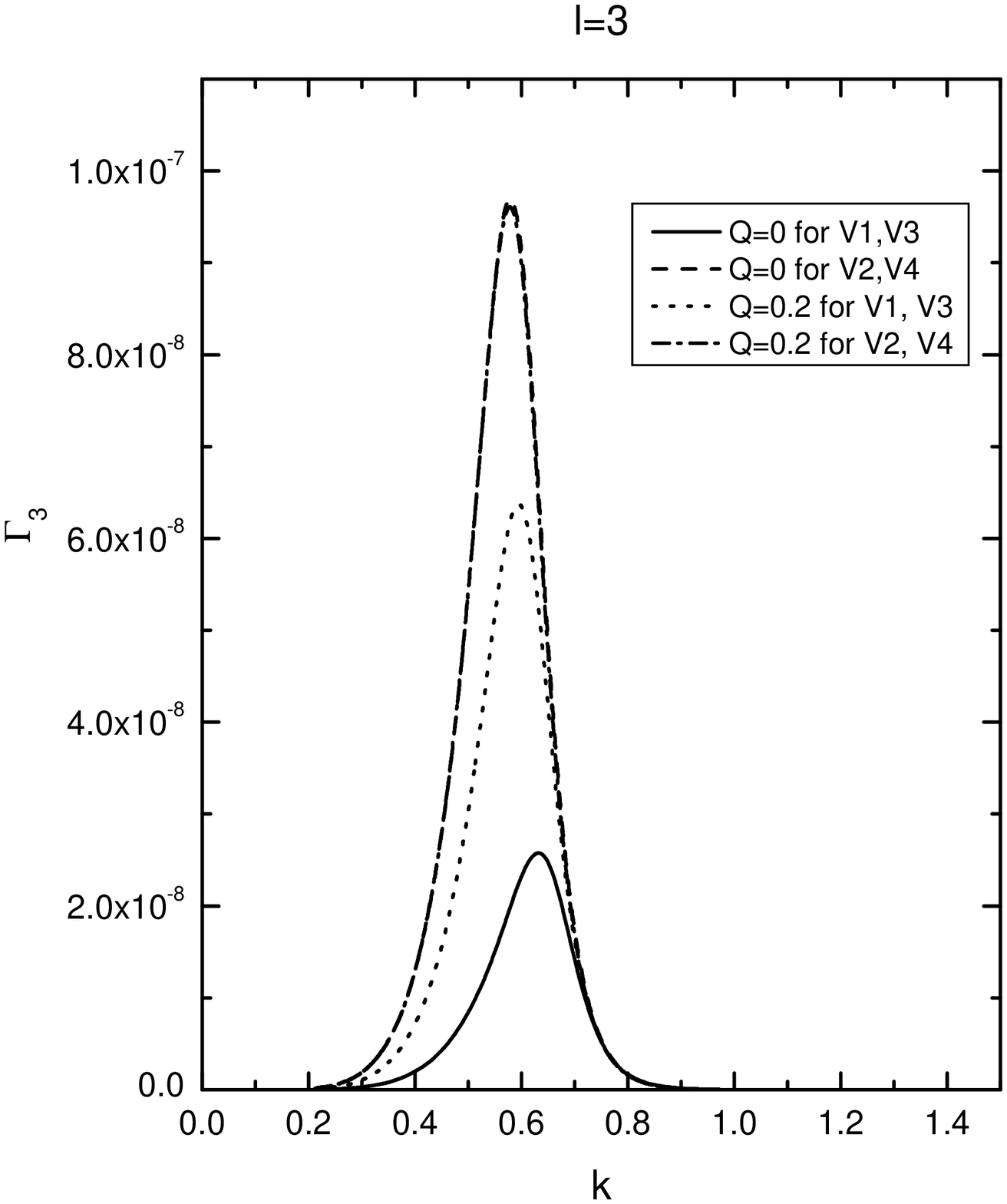}}
\nonumber
\end{minipage} \nonumber
\caption{{The partial emission spectrum $\Gamma_l$ with $M=1$ by
taking $l=2$ and $l=3$ in the Schwarzschild black hole and the
brane-world black hole.}}\label{f7}
\end{figure}

Fig.\ref{f7} illustrates the partial emission spectrum. The peak
values of the emission spectrum at $l=2$ are much larger than those
at $l=3$, approximately by $\sim 10^2$ times. Therefore the main
emission spectrum for the gravitational perturbation can be
represented by that of $l=2$ in Fig.\ref{f7} due to the suppression
for the higher $l$ modes. But the total emission spectrum of the
brane-localized gravitons is not well defined here, because the high
$l$ modes easily break down the constraint $Q < \frac{9M^2}{8n}(l
\ne 0)$, where the flux on the brane is no longer conserved but with
the leakage to the bulk which makes our brane localized
investigation not appropriate. As in the case of scalar field, there
also exist critical values $Q_h$ for the potential $V_2$(or $V_4$).
If $ Q_h < Q < \frac{{9M^2 }}{{8n}}$, the peak of the emission
spectrum is not beyond that in Schwarzschild black hole case and if
$Q < Q_h$, the emission spectrum of the brane-world black hole is
stronger than that of the 4D Schwarzschild case despite of its lower
temperature. $Q_h \approx 0.48$ for $l=2$ and $Q_h \approx 0.19$ for
$l=3$. But this critical value has not been observed in the Hawking
radiation for $V_1$(or $V_3$) where the emission of the brane world
black hole is always stronger than that of the 4D Schwarzschild
black hole due to the emission of the brane-localized gravitons
although it has lower temperature than that of the Schwarzschild
hole. The mathematical reason is that the increase of the absorption
cross section is not big enough to compensate the change of
temperature in the denominator of Eq.(\ref{e24}) for $V_2$(or
$V_4$), while $V_1$(or $V_3$) creates the large enhancement of the
absorption which makes the numerator win the competition against the
denominator. This unusual phenomenon can be considered as the pure
effects of the bulk influence, which is hard to be imagined in 4D
black holes. We have calculated the partial emission rate $H_l$ by
integrating Eq.(\ref{e24}) in the Schwarzschild black hole with
$M=1$ and also for the brane-world black hole with the same mass and
$Q=0.2$. The results are shown the results in Table 3. The emission
rates arise almost two times by the tidal parameter $Q=0.2$ for
$V_1$(or $V_3$) but do not change much for $V_2$(or $V_4$).

\vspace{0.5cm}
\begin{center}

\begin{tabular}{l|l|l|l|r} \hline
$ $           &   $l=2$   &     $l=3$       \\  \hline \hline
Sch. $H_l$ for $V_1$, $V_3$   &   $3.3 \times 10^{-7}$   &  $5.0 \times 10^{-9}$       \\
Sch. $H_l$ for $V_2$, $V_4$   &   $1.9 \times 10^{-6}$   &  $1.8 \times 10^{-8}$       \\
Tdl. $H_l$ for $V_1$, $V_3$      &   $5.5 \times 10^{-7}$     &  $1.2 \times 10^{-8}$   \\
Tdl. $H_l$ for $V_2$, $V_4$      &   $2.0 \times 10^{-6}$     &  $1.8 \times 10^{-8}$   \\
\hline
\end{tabular}

\vspace{0.1cm} \large{Table III}

\vspace{0.1cm} {\it The partial emission rate $H_l$ in the
Schwarzchild and tidal charge black hole backgrounds.}
\end{center}
\vspace{0.5cm}

Now we discuss briefly the case for $ Q > \frac{{9M^2 }}{{8n}}\quad
(l\ne 0)$. The non-vanishing imaginary part of the potentials $V_i$
leads to the non-zero r.h.s of the Wronskian equation
\begin{equation} \label{e39}
f^2 W' + ff'W = 2iZ_i^* Z_i {\mathop{\rm Im}\nolimits} [V_i ] \equiv
iJ_I,
\end{equation}
whose solution is easily obtained
\begin{equation} \label{e40}
W = \frac{{iC}}{f} + \frac{i}{f}\int {\frac{{J_I }}{f}dr}.
\end{equation}
If we still define the 'current density' as in \cite{sanchez}, the
'flux' conservation on the brane can not be held due to the
existence of the second term in the r.h.s of Eq.(\ref{e40}). This
means that for a certain tidal charge parameter, the 'flux' of the
gravitational perturbation in the higher modes satisfying $ Q >
\frac{{9M^2 }}{{8n}}\quad (l\ne 0)$ cannot stay on the brane but
leave into the bulk inevitably (at least part of it does so), though
the perturbation is created on the brane. In other words, when the
tidal parameter $Q  > 9/16$, the gravitational perturbation may
leave the brane and could be very hard to detect even for the low
angular index $l=2$ in our method. This is an effect of the extra
dimension. The detailed and precise analysis on the absorption and
emission of the more negatively tidal charged brane-world black hole
is not clear at this moment, since the exact solution of the metric
in the 5D bulk is still unknown. On the other hand, many researches
\cite{s10} show that the emission is denominated by that on the
brane, hence the emission of the non brane-localized gravitons is
beyond our discussion and cannot be analyzed appropriately in our
present frame.

\section{Conclusion and Discussion}
In this paper, we have studied the perturbations around the
brane-world black hole whose projected bulk effects are represented
by the tidal charge. The positive tidal charged black hole is of the
RN type and the negative one has only one horizon, which is larger
than that of the 4D Schwarzschild case surrounding the central
singularity. The effect of the negative tidal charge strengthens the
gravitational field and lowers the temperature of the black hole
\cite{Tdl}.

We have calculated the QNMs for both massless scalar field
perturbation and gravitational perturbations. We have compared them
to the case of the 4D Schwarzschild and RN black holes with the same
mass. For the negative tidal charge case, unlike those observed in
RN black hole, there is no critical charge where QNM differs below
and above this critical charge
\cite{star}\cite{wangb1}\cite{wangb2}. When the negative tidal
charge becomes more negative, the perturbations stay longer with
less oscillations regardless of what types of perturbations. The
QNMs of the negative tidal charge and positive tidal charge black
holes are separated by that of the 4D Schwarzschild black hole.

However, the influence of the tidal charge on the Hawking radiation
are not as simple as that on the QNMs. In the scalar field, the
negative tidal charge $Q$ suppresses the effective potential barrier
and the absorption cross section increases correspondingly as a
consequence of its effect strengthening the gravitational field. It
is interesting that the emission spectrum does not simply decrease
with the temperature as that in the background of RN black hole
\cite{s11}. There exist the critical values $Q_h$. When $Q > Q_h$,
the peak of the emission spectrum is smaller than that of
Schwarzschild black hole of the same mass. But if we take $Q < Q_h$,
despite that the temperature is lower, the emission becomes stronger
compared to the Schwarzschild black hole. This is the result of the
enhancement of the absorption cross section and the decrease of the
temperature according to Eq.(\ref{e24}) caused by the negative tidal
charge, which changes our usual picture about the Hawking radiation
in 4D black holes. We emphasis that it is a new behavior which has
not been observed before.

As for the graviton, we find that when $ Q > \frac{{9M^2
}}{{8n}}\quad (l\ne 0)$, the conservation of the graviton flux on
the brane is broken and the leakage of the gravitons requires the
exact solution of the brane world black hole in the bulk which has
not been known yet. We have investigated only the Hawking radiation
of the brane-localized gravitons under the condition $ Q <
\frac{{9M^2 }}{{8n}}\quad (l\ne 0)$. As pointed out before, there
are two kinds of gravitational perturbations: $V_1$(or $V_3$)
referring to those generated mainly by the fluctuation of the bulk
effects and $V_2$(or $V_4$) to those created by the factors on the
brane. All these potential barriers are suppressed by the negative
tidal charge. As a result, the absorption cross sections increase.
For the potential $V_2$(or $V_4$), the behaviors of the partial
emission spectrums are much similar to those in the scalar field and
there is a critical value $Q_h$, beyond which the peak of the
emission spectrum is smaller and below this $Q_h$, the emission
spectrum is stronger compared to that of the 4D Schwarzschild black
hole, although with the negative tidal charge we always have lower
black hole temperature. But for the potential $V_1$(or $V_3$), no
matter what is the value of the tidal charge, the emission spectrum
of the brane world black hole is always stronger than that of the Schwarzschild
black hole for the brane-localized gravitons. The phenomenon of
increasing emission spectrum with decreasing temperature in the
brane world black hole has not been observed before. This is an
effect of the negative tidal charge caused by the bulk effect.

We conclude that both the QNMs and the Hawking radiation have given
signatures of negative tidal charge due to the bulk effects in the
brane-world black hole, which differs from that of the 4D
Schwarzschild black hole. We expect that these signatures can be
observed in the future experiments, which could help us learn the
properties of the extra dimensions.

\begin{acknowledgments}
This work was partially supported by  NNSF of China, Ministry of Education
of China and Shanghai Science and Technology Commission. We would like to acknowledge E. Abdalla and E. Papantonopoulos for helpful discussions.
\end{acknowledgments}

\end{document}